\documentclass[twocolumn]{aastex62}
\usepackage{epsfig}
\usepackage{natbib}
\usepackage{color}
\usepackage[squaren]{SIunits}
\submitjournal{Astronomical Journal} 
\accepted{July 24th, 2018}

\newcommand\gaia{{\em Gaia }}

\shortauthors{Donor et al.}
\shorttitle{OCCAM: II. Precision Abundances with APOGEE DR14}

\begin{document}

\title{The Open Cluster Chemical Abundances and Mapping Survey: II.\\ 
  Precision Cluster Abundances for APOGEE using SDSS DR14}

\correspondingauthor{John Donor}
\email{j.donor@tcu.edu}

\author{John Donor}
\affiliation{Department of Physics \& Astronomy, Texas Christian University,
TCU Box 298840, \\Fort Worth, TX 76129, USA (j.donor, p.frinchaboy, b.a.thompson1,
 j.oconnell, b.r.meyer@tcu.edu)}
\author[0000-0002-0740-8346]{Peter M. Frinchaboy}
\affiliation{Department of Physics \& Astronomy, Texas Christian University,
TCU Box 298840, \\Fort Worth, TX 76129, USA (j.donor, p.frinchaboy, b.a.thompson1,
 j.oconnell, b.r.meyer@tcu.edu)}
\author{Katia Cunha}
\affiliation{Observat{\'o}rio Nacional, S{\~a}o Crist{\'o}v{\~a}o, Rio de Janeiro, Brazil}
\author{Benjamin Thompson}
\affiliation{Department of Physics \& Astronomy, Texas Christian University,
TCU Box 298840, \\Fort Worth, TX 76129, USA (j.donor, p.frinchaboy, b.a.thompson1,
 j.oconnell, b.r.meyer@tcu.edu)}
\affiliation{GitHub, Inc., 88 Colin P. Kelly Street, San Francisco, CA 94107.}
\author{Julia O'Connell}
\affiliation{Department of Physics \& Astronomy, Texas Christian University,
TCU Box 298840, \\Fort Worth, TX 76129, USA (j.donor, p.frinchaboy, b.a.thompson1,
 j.oconnell, b.r.meyer@tcu.edu)}
\author{Gail Zasowski}
\affiliation{Department of Physics \& Astronomy, University of Utah, 115 S. 1400 E., Salt Lake City, UT 84112, USA}
\author{Kelly M. Jackson}
\affiliation{Department of Physics \& Astronomy, Texas Christian University,
TCU Box 298840, \\Fort Worth, TX 76129, USA (j.donor, p.frinchaboy, b.a.thompson1,
 j.oconnell, b.r.meyer@tcu.edu)}
\affiliation{Current Address: Texas Instrument Corporation, Fort
  Worth, TX }
\author{Brianne Meyer McGrath}
\affiliation{Department of Physics \& Astronomy, Texas Christian University,
TCU Box 298840, \\Fort Worth, TX 76129, USA (j.donor, p.frinchaboy, b.a.thompson1,
 j.oconnell, b.r.meyer@tcu.edu)}
\affiliation{Department of Physics, University of Colorado Colorado Springs,\\
1420 Austin Bluffs Pkwy, Colorado Springs, CO USA 80918}
\author{Andr{\'e}s Almeida}
\affiliation{Instituto de Investigaci{\'o}n Multidisciplinario en Ciencia y Tecnolog{\'i}a,\\ Universidad de La Serena, Benavente 980, La Serena, Chile (aalmeida@userena.cl)}
\author{Dmitry Bizyaev}
\affiliation{Apache Point Observatory and New Mexico State
    University, P.O. Box 59, Sunspot, NM, 88349-0059, USA(kpan, dmbiz@apo.nmsu.edu)}
\affiliation{Sternberg Astronomical Institute, Moscow State
    University, Moscow, Russia}
\author{Ricardo Carrera}
\affiliation{Astronomical Observatory of Padova, National Institute of Astrophysics, \\Vicolo Osservatorio 5 - 35122 - Padova (ricardo.carrera@oapd.inaf.it)}
\author{D. A. Garc{\'i}a-Hern{\'a}ndez}
\affiliation{Instituto de Astrofısica de Canarias,
38205 La Laguna, \\Tenerife, Spain (ozamora, agarcia@iac.es)}
\affiliation{Universidad de La Laguna (ULL), Departamento de Astrofísica, E-38206 La 
Laguna, \\Tenerife, Spain}
\author{Christian Nitschelm}
\affiliation{Unidad de Astronom{\'i}a, Universidad de Antofagasta, Avenida Angamos 601, \\Antofagasta 1270300, Chile (christian.nitschelm@uantof.cl)}
\author{Kaike Pan}
\affiliation{Apache Point Observatory and New Mexico State
    University, P.O. Box 59, Sunspot, NM, 88349-0059, USA(kpan, dmbiz@apo.nmsu.edu)}
\author{Olga Zamora}
\affiliation{Instituto de Astrofısica de Canarias,
38205 La Laguna, \\Tenerife, Spain (ozamora, agarcia@iac.es)}
\affiliation{Universidad de La Laguna (ULL), Departamento de Astrofísica, E-38206 La 
Laguna, \\Tenerife, Spain}


\begin{abstract}
The Open Cluster Chemical Abundances and Mapping (OCCAM) survey 
aims to produce a comprehensive, uniform, infrared-based spectroscopic dataset for
hundreds of open clusters, and to constrain key Galactic dynamical
and chemical parameters from this sample.
This second contribution from the OCCAM survey 
presents analysis of 259 member stars with [Fe/H] determinations in
19 open clusters, using Sloan Digital Sky Survey Data Release 14 (SDSS/DR14) data from 
the Apache Point Observatory Galactic Evolution Experiment
(APOGEE) and ESA {\em Gaia}.  This analysis, which includes clusters with $R_{GC}$ ranging from 7 to 13 kpc, measures an [Fe/H] gradient of $-0.061 \pm 0.004$ dex
kpc$^{-1}$.
We also confirm evidence of a significant positive 
gradient in the $\alpha$-elements ([O/Fe], [Mg/Fe], and [Si/Fe]) 
and present evidence for a  significant negative gradient in 
iron-peak elements ([Mn/Fe] and [Ni/Fe]).

\end{abstract}


\vspace{0.38in}
\section{ INTRODUCTION }

The detailed chemical evolution of stars in galaxies provides key
information on how galaxies grow and evolve.  
Star clusters provide an age-datable
tracer for measuring the growth and evolution of the Galaxy.
One key measurement is the disk radial chemical trend or gradient as traced by open clusters \citep{janes_79}.
The Galactic abundance gradient has
been fit by a single linear gradient (e.g. \citealt{friel_93, friel_95, carraro_98, friel_02}), and increasingly more commonly by 
a 2-function gradient (e.g. \citealt{bragaglia2008, sestito2008, friel2010, carrera_2011,reddy_16}).

Recent work using open clusters has consistently found a metallicity gradient (d[Fe/H]/dR$_{GC}$) between roughly $-$0.05 dex kpc$^{-1}$ (\citealt{reddy_16}, hereafter R16) and  $-$0.09 dex kpc$^{-1}$ \citep{yong_2012, friel_95, carraro_98} for clusters between 6 kpc $< R_{GC} < 14$ kpc. Others have reported qualitatively similar trends, but do not quote a metallicity gradient measurement \citep{donati_2015, magrini_2015, magrini_2017,Casamiquela_2017}.
But while the general negative shape of the abundance trend is well agreed upon, no consensus has been reached on the steepness of the gradient.
\citet{carrera_2011} shed some light on this discrepancy by showing the difference between a gradient measured to R$_{GC} =12.5$ kpc ($-$0.070  $\pm$ 0.005)
 and all the way to R$_{GC} =25$ kpc ($-$0.046  $\pm$ 0.010); \citet{frinchaboy_13} show a similar discrepancy using [M/H]. 

Another unavoidable problem that has made this measurement difficult is systematic offsets between studies of chemical abundance and distance. This inevitably introduces some systematic uncertainties when a compilation of results from the literature is used. { Recent work has sought to correct for systematic uncertainties uncertainties by ``homogonizing" their samples. R16 take equivalent width measurements from the literature, but use a uniform line list for their analysis. \citep{netopil} homogenized a large photometric sample using a literature compilation of high-resolution spectroscopic studies, however they do not homogenize the spectroscopic studies.}


This paper presents an important contribution to the field by utilizing a homogeneous spectroscopic data set, with all stars observed by the same telescope and analyzed with the same abundance analysis pipeline: the Apache Point Observatory Galactic Evolution Experiment (APOGEE; \citealt{apogee}). 
We present a high reliability sample of stars which are open cluster members, bulk cluster parameters, and Galactic abundance gradient using the APOGEE/OCCAM DR14 sample. 
This paper is organized as follows: 
The OCCAM target selection is presented in \S2 and cluster membership analysis in \S3.  We compare our work to other studies in \S4.  Finally, in \S5 we present our findings on the Galactic metallicity gradients using APOGEE-based abundances, and in \S6 we consider gradients in other elements.  

\section{OCCAM Target Selection \label{sec:2MASS}}

The OCCAM target selection includes targets selected in two different ways. First we selected known members of a subset of open clusters that were observed for calibration purposes.  Stars with previous abundance determinations \citep[e.g.,][]{cohen_1980,origlia_2006,Carretta_2007,carraro_2006,bragaglia_2001,yong_2005, M67_Taut2000, pancino_2010, basu_2011, smith_1987}
and/or high quality RV-based membership studies  \citep[e.g.,][]{hole_2009, geller_2008, geller_2010, mermilliod_2008} were targeted.
These calibration cluster targets can be identified in DR14 though a specific targeting flag \citep[{\em apogee\_target2} = 10 and/or {\em apogee2\_target2} = 10; ][]{zasowski13, zasowski17}

The second method selected ``likely" cluster targets based upon their location in the cluster color-magnitude diagram (CMD) using the colors in the surveys 2MASS \citep{2MASS_DR_allsky} and WISE \citep{WISE_mission_description}.

\begin{figure*}
	\begin{center}
		\vskip-0.3in
        \epsscale{1.2}
    \plotone{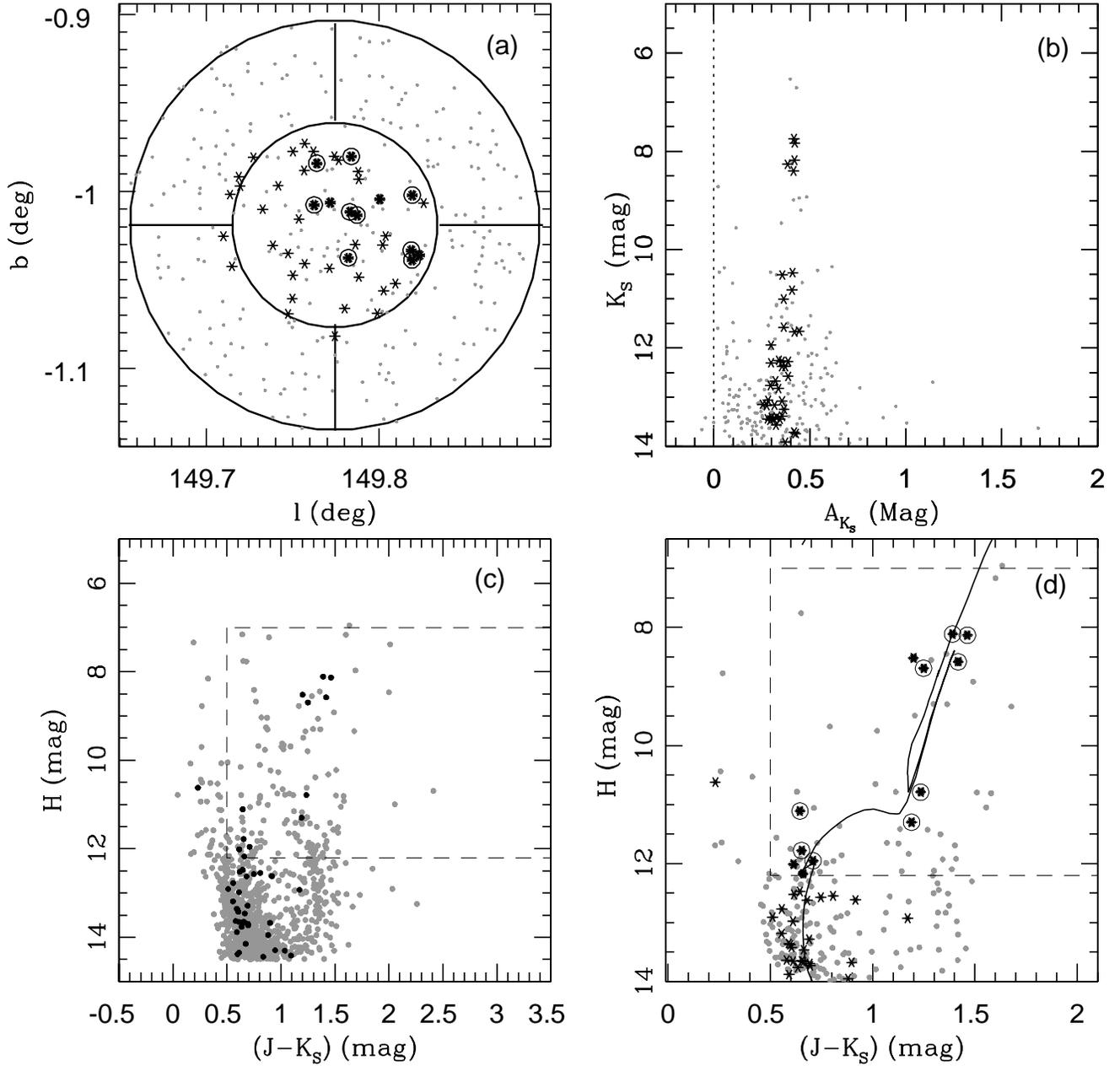}
	\end{center}
	\caption{ \small Sample analysis for the cluster King 7 utilizing
		2MASS+WISE data (\citealt{AP_Target_selection}). a) Galactic latitude and longitude  for all stars (gray) within the $2R_{cl}$ area to
		be analyzed, stars selected to be likely members from the photometry
		extinction analysis are shown in black.  Prime APOGEE targets are
		circled. b) Distribution of $A_{K_s}$ for all stars in the NGC 6802 
		sample area, black points denote stars with $1.1 R_{cl}$ within the
		determined mean cluster $A_{K_s}$ range.  c) Color-magnitude diagram
		(CMD) for all stars in the analysis area (gray). The dashed box
		denotes the SDSS-III/APOGEE target selection region.  Black points
		denote stars selected as likely members from their $A_{K_s}$. d) CMD of only likely
		cluster members overplotted with the Padova Isochrone
		(Marigo et al.\ 2008) using the
		clusters parameters from \cite{dias_catalog}.  Circled stars
		denote identified high-probability stars for APOGEE target selection
		(also see the sky distribution (a)).
	}
	\label{fig:cl}    
\end{figure*}

The combination of 2MASS and {\em WISE} 
photometry allows for a direct assessment of the line-of-sight reddening to any particular star. 
The long wavelength regime of spectral energy distributions (SEDs) of stars have the same Rayleigh-Jeans shape, equivalent to saying that the Vega-based, {\it intrinsic} colors of all stars are nearly constant for the correct combination of filters, as seen in Figure 5 of \citet{h_4.5m}. Thus, the {\it observed} mid-IR colors contain information on the reddening to a star {\it explicitly}, whereas the NIR SEDs contain information on the stellar types.  

By assuming constancy of the intrinsic stellar ($H$$-$$4.5\mu$m) colors in the Rayleigh-Jeans regime, $E(H$$-$$4.5\mu$m) is derived directly from the observed ($H$$-$$4.5\mu$m) color \citep{h_4.5m}. The spread from different populations, RGB, red clump and main sequence, is minimized for this combination yielding an intrinsic spread of less than 0.09 mag in color for all but the reddest and bluest stars.
Since the primary purpose in using this technique is to ``clean'' the cluster from the field, small systematics are not a concern.  Also, the reddest main sequence stars that would belong to a cluster are too faint for these surveys.


\cite{AP_Target_selection} devised a technique to utilize the extinction ($A_{K_S}$) derived from the RJCE technique 
to distinguish and isolate cluster stars from foreground and background contamination. This technique consists of isolating a region of approximately twice the cluster's catalog radius ($R_{Dias}$; \citealt{dias_catalog}) and dividing it into five regions (see Figure~\ref{fig:cl}a).  We utilize four ``field'' regions and the cluster region (radius = $R_{Dias}$). The field is divided in order to account for dust clouds
and any other source of background variability.

We subtract the median area-scaled ``field" star density from the ``cluster'' star numbers within a given $A_{K_s}$ range, and scan this range across all available $A_{K_s}$ values that have at least 15 stars (see Figure~\ref{fig:cl}b). The window of extinction with the highest concentration of stars within the inner radius will reveal the cluster (Figure~\ref{fig:cl}c \& d). We then optimize the cluster isolation surveying a grid of $A_{K_s}$ width, $A_{K_s}$ stepsize, and allowed $\sigma_{A_{K_s}}$ values. 

We present a demonstration of this technique utilizing the cluster King 7, shown in Figure~\ref{fig:cl}. Figure~\ref{fig:cl}a first shows the area explored by our analysis in Galactic latitude and longitude. As described above, we selected likely cluster members utilizing the $A_{K_s}$ as shown in Figure~\ref{fig:cl}b.  For King 7 we find a low, but non-negligible extinction or reddening to the cluster. A CMD of this cluster (Figure~\ref{fig:cl}c) is generated which highlights the member stars with $A_{K_s}$ values within the selected window of extinction, where the dashed box denotes the area where the  SDSS/APOGEE project selects targets ($8.0 < H < 12.2$ and $J-K_S \ge 0.5$). Finally, we compare our ``cleaned'' cluster CMD to the Padova isochrone \citep{padova_isos} utilizing catalog values \citep{dias_catalog} for King 7 and find a good match.  By comparing the CMD with isochrone values, when available, we are able to isolate candidate open cluster stars with a high probability for membership.  
The APOGEE project requires this cleaning for most clusters for {three} reasons: 1) most open clusters are found at low Galactic latitude and thereby are heavily contaminated with field stars. 2) Due to the large SDSS telescope field of view \citep{sloan_telescope},
the minimum fiber-to-fiber distance is fairly large ($\ge 1$ arcmin), which only allows for the targeting of a handful of stars ($\sim 5-10$) per cluster for the most poorly studied, distant, and reddened clusters. {3) Prior to {\it Gaia}  the proper motion data required for high-quality reliable membership determinations were only available for a few clusters.}

Likely open cluster members selected by this method are identified in DR14 \citep{dr14}, and previous DRs (10,12,13) through specific targeting flag \citep[{\em apogee\_target1} = 9 and/or {\em apogee2\_target1} = 9; ][]{zasowski13, zasowski17}\footnote{We did not limit our analysis to stars with just these targeting flags. Random field stars and additional specific targeted cluster programs, and other calibration flags may also apply to the targets analyzed here.}.

\section{Analysis}

\subsection{OCCAM Observed Stars in SDSS4 DR14}

The primary spectroscopic
data for OCCAM comes from the Apache Point Observatory Galactic Evolution Experiment (APOGEE; \citealt{apogee}), which is part of the Sloan Digital Sky Survey-III and IV surveys (SDSS; \citealt{sloan3_overview, sdss4}), utilizing the $2.5\ \meter$ Sloan Foundation telescope \citep{sloan_telescope} at Apache Point Observatory. APOGEE is a near-infrared ($1.514\ \micro \meter $ to $1.696\ \micro \meter $) spectroscopic survey, primarily focusing on the galactic disk \citep{zasowski13,zasowski17}. The survey uses multi-fiber spectrographs
\citep{wilson_2012}, allowing for simultaneous observations of 300 stars. 

The APOGEE data reduction pipeline \citep{nidever_2015, holtzman_2015} provides high precision radial velocities (RVs).
Stellar parameters (T$_{eff}$ , $\log g$, [M/H], [C/M], [N/M], [$\alpha$/M]), and
detailed abundances for individual elements, such as, Fe, C, N, O,
Al, Si, Ca, Ni, Na, S, Ti, Mn, K, and Cu, are derived automatically by the ASPCAP pipeline \citep{aspcap}. The APOGEE survey provides the {\it uniform} chemical data that underpin this study of open cluster members.

The targets selected for analysis were observed from August 2011 to July 2014 (APOGEE-1), and from July 2014 to July 2016 (APOGEE-2). These data were released as part of the 14th Data Release of SDSS (DR14; \citealt{dr14}), which included APOGEE data for over 250,000 stars. All APOGEE data, from the beginning of APOGEE-1, were reduced using the latest data reduction pipeline (full description of this pipeline is presented in Holtzman et al. 2018, {\it submitted} ).

For this study, we analyzed all stars within 2$\times$ the cluster radius \citep{mwsc_catalog} for 19 clusters that resulted in a sample of 1361 stars. This entire sample is listed in Table \ref{tab:full_sample_stars} for reference, along with our final membership probabilities and a classification for each star (\S \ref{sec:mem_criteria}).


\subsection{OCCAM 
Membership Criteria}\label{sec:mem_criteria}

Using the stellar radial velocities and derived metallicities as initial discriminators,  
APOGEE data alone can provide a first guess at cluster membership based on the ``bulk" RV and [Fe/H] for the cluster region on the sky and comparing each star to the average values. We then further constrain the membership using proper motions measured by \gaia \citep{gaia_mission,gaia_dr2,gaia_astrometry}.

\subsubsection{Quantifying Membership Probability}\label{sec:memAnalysis}

\begin{figure*}
	\begin{center}
		\vskip-0.3in
        \epsscale{0.9}
    \plotone{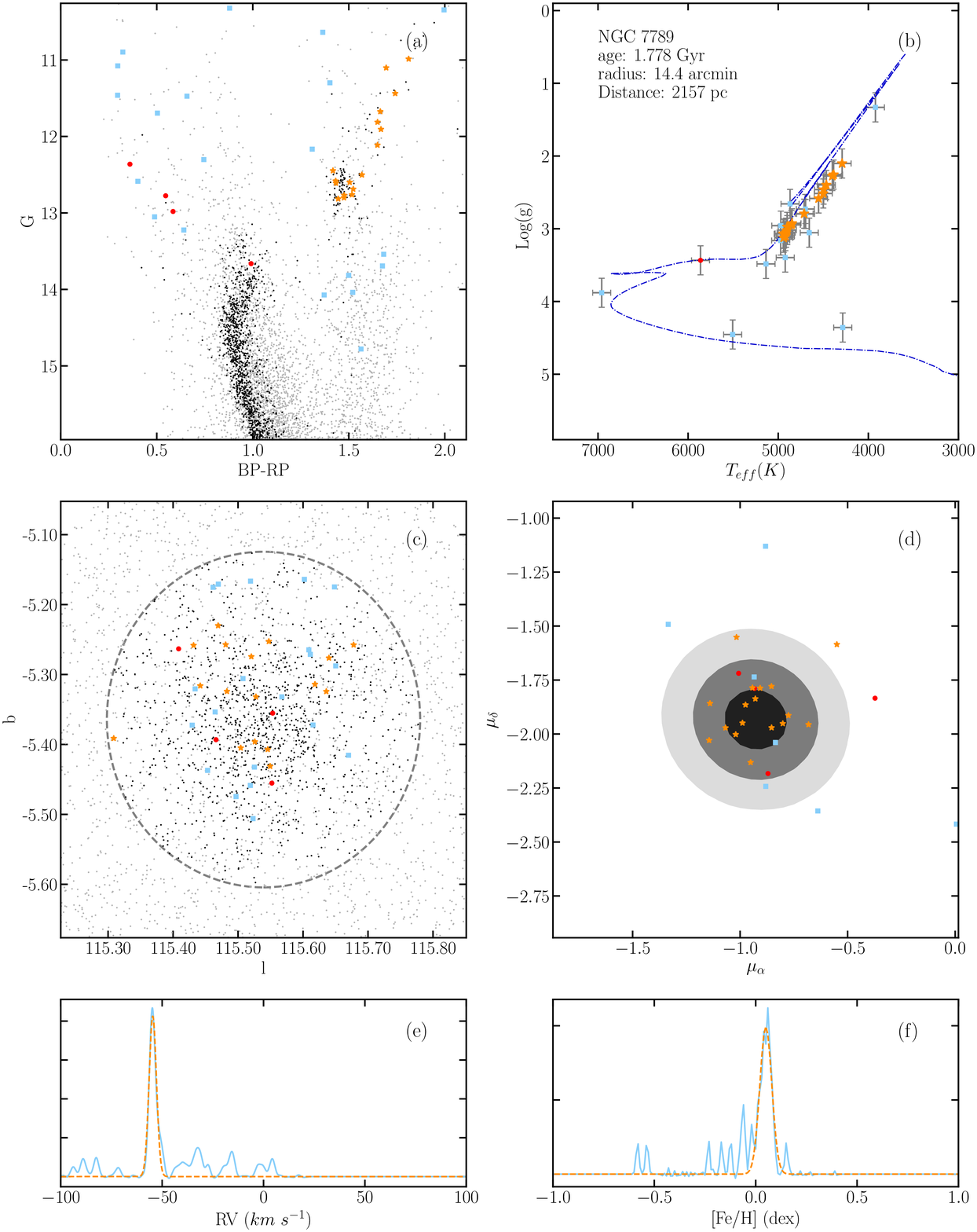}
	\end{center}
	\caption{ \small A summary of the membership analysis for open cluster NGC 7789. The complete figure set (19 images) is available in the online journal.
    (a) The \gaia CMD \citep{gaia_dr2,gaia_photometry}, with proper motion members shown in black, others in gray. Likely APOGEE members are shown as orange stars, non-members as blue squares, and APOGEE stars passing an RV and proper motion membership cut, but failing a $\log(g)$ cut for metallicity reliability are shown as red circles. 
    (b) The $T_{eff}-\log(g)$ diagram for the cluster with an isochrone based on MWSC Catalog ages \citep{mwsc_catalog} shown for reference. Error bars shown are characteristic, and a possible global offset in $\log(g)$ is seen. 
    (c) The cluster area on the sky.
    (d) A contour plot of the 2D Gaussian fit to the kernel smoothed proper motions. Contours show $1 \sigma$ intervals.
    (e) The Gaussian kernel density convolution in RV, with a Gaussian fit shown in orange. 
    (f) The Gaussian kernel density convolution in [Fe/H], with a Gaussian fit shown in orange.}
	\label{fig:ngc6819}
\end{figure*}

The ``bulk" behavior is found by convolving all measurements using a Gaussian kernel smoothing routine{, based on the methods from \citet{frinchaboy_2008}}. The first analysis is in RV. In order to distinguish the cluster from field stars, 2 samples are computed: stars within 2 cluster radii (from \citealt{mwsc_catalog}, except where we enforced a minimum radius of 5 arcminutes) of the cluster center, and stars between 1 and 2 cluster radii of the center. 
The results from the ``outer" stars are subtracted from the ``total" result, leaving a peak where the cluster stars fall, as seen in Figure \ref{fig:ngc6819}e (in blue).

\begin{deluxetable*}{lrrrrrrccc}[h!]
\tablecaption{OCCAM sample from APOGEE data used for membership analysis\tablenotemark{a}  \label{tab:full_sample_stars}}
	\tablehead{
    \colhead{Cluster} & 
    \colhead{2MASS ID} &
    \colhead{RV} & 
    \colhead{[Fe/H]} &
    \colhead{$\mu_{\alpha}$\tablenotemark{b}} &
    \colhead{$\mu_{\delta}$\tablenotemark{b}} &
    \colhead{RV } & 
    \colhead{[Fe/H] \tablenotemark{c}}& 
    \colhead{PM} &
    \colhead{Member\tablenotemark{d}}\\[-2ex]
    \colhead{name} & \colhead{} &
    \colhead{(km s$^{-1}$)} & 
    \colhead{(dex)} & 
    \colhead{(mas yr$^{-1}$)} &
    \colhead{(mas yr$^{-1}$)} &
    \colhead{Prob} & 
    \colhead{Prob} &
    \colhead{Prob} &
    \colhead{}
    }
	\startdata
NGC 6819 & 2M19401402+4016306 &$-$56.4 $\pm$ 0.0 & $+$0.06 $\pm$ 0.01 &$-$5.85 $\pm$ 0.03 & $-$9.04 $\pm$ 0.03 & 0.00 & \phs0.00 & 0.00 & NM\\
NGC 6819 & 2M19401466+4004598 &$-$18.9 $\pm$ 0.0 & $-$0.03 $\pm$ 0.01 &$+$50.96 $\pm$ 0.04 & $+$99.82 $\pm$ 0.04 & 0.00 & $-$1.00 & 0.00 & NM\\
NGC 6819 & 2M19401937+4015495 &$-$52.3 $\pm$ 0.2 & $-$0.35 $\pm$ 0.01 &$-$5.93 $\pm$ 0.03 & $-$6.26 $\pm$ 0.03 & 0.00 & \phs0.00 & 0.00 & NM\\
NGC 6819 & 2M19402284+4006008 &$-$50.3 $\pm$ 0.1 & $-$0.34 $\pm$ 0.01 &$-$1.60 $\pm$ 0.03 & $-$2.91 $\pm$ 0.03 & 0.00 & \phs0.00 & 0.00 & NM\\
NGC 6819 & 2M19403569+4005038 &$+$2.7 $\pm$ 0.1 & $-$0.45 $\pm$ 0.01 &$-$3.76 $\pm$ 0.03 & $-$3.46 $\pm$ 0.03 & 0.00 & \phs0.00 & 0.00 & NM\\
NGC 6819 & 2M19403684+4015172 &$+$2.1 $\pm$ 0.0 & $+$0.15 $\pm$ 0.01 &$-$3.00 $\pm$ 0.03 & $-$3.66 $\pm$ 0.03 & 0.00 & \phs0.00 & 0.54 & NM\\
NGC 6819 & 2M19404262+4003043 &$-$35.2 $\pm$ 0.1 & $+$0.07 $\pm$ 0.01 &$-$4.36 $\pm$ 0.04 & $-$19.13 $\pm$ 0.04 & 0.00 & \phs0.00 & 0.00 & NM\\
NGC 6819 & 2M19404341+4020235 &$-$11.8 $\pm$ 16.3 & $-$0.01 $\pm$ 0.01 &$+$0.65 $\pm$ 0.09 & $-$3.89 $\pm$ 0.09 & 0.00 & $-$1.00 & 0.00 & NM\\
NGC 6819 & 2M19404803+4008085 &$+$2.4 $\pm$ 0.1 & $+$0.09 $\pm$ 0.01 &$-$2.98 $\pm$ 0.04 & $-$3.83 $\pm$ 0.04 & 1.00 & \phs0.82 & 0.94 & GM\\
NGC 6819 & 2M19404965+4014313 &$+$3.1 $\pm$ 0.0 & $+$0.12 $\pm$ 0.01 &$-$2.84 $\pm$ 0.03 & $-$3.72 $\pm$ 0.03 & 0.96 & \phs0.94 & 0.70 & GM\\
NGC 6819 & 2M19405020+4013109 &$+$4.3 $\pm$ 0.0 & $+$0.15 $\pm$ 0.01 &$-$3.03 $\pm$ 0.04 & $-$3.84 $\pm$ 0.04 & 0.67 & \phs0.59 & 0.85 & GM\\
    \multicolumn{10}{c}{ \nodata }\\
\enddata
\tablenotetext{a}{Table \ref{tab:full_sample_stars} is published in its entirety in the
  electronic edition of Astronomical Journal, A portion is shown here for guidance regarding its form and content.}
\tablenotetext{b}{From \gaia DR2 \citep{gaia_dr2,gaia_astrometry}}
\tablenotetext{c}{A [Fe/H] membership probability of -1 corresponds to a check against $\log(g)$ for [Fe/H] reliability. Stars which failed may be members, but are flagged because [Fe/H] is not a reliable membership discriminator for these stars.}
\tablenotetext{d}{Possibilities here are: GM (giant member), DM (dwarf member), and NM (non-member). We differentiate between giants and dwarfs due the $\log(g)$ cut mentioned previously: these dwarfs may be members, but their metallicities may not be reliable.}
\end{deluxetable*}

The process is repeated for [Fe/H], seen in Figure \ref{fig:ngc6819}f (in blue) this time subtracting the stars farther than $3\sigma$ from the cluster RV previously identified from the whole field ($\sigma$ is small in practice, thus $3\sigma$ is appropriate for keeping cluster stars without including field stars incidentally close in RV space). If there are at least two APOGEE stars that are cluster members, the smoothing routine will leave behind a larger peak where their values combined. The shape is approximately Gaussian, so a Gaussian profile is fit for both RV and [Fe/H]. When normalized, this Gaussian fit can be used as a membership probability distribution in RV or [Fe/H] space, seen in Figure \ref{fig:ngc6819}e \& f (overlaid in orange).


{Using proper motion data from \gaia DR2 \\ \citep{gaia_mission,gaia_dr2,gaia_astrometry},} a 2-dimensional Gaussian smoothing routine is applied in proper motion space. Again, 2 samples are computed: all stars within twice the cluster radius and stars outside the cluster's radius, then the outside sample is subtracted from the full sample. A 2D Gaussian is fit to the remaining peak and membership probabilities are assigned, shown in Figure \ref{fig:ngc6819}d.

Finally, a $3 \sigma$ criterion is adopted for likely membership: a star with parameters falling within $3 \sigma$ of the cluster mean in  [Fe/H], RV, and proper motion is considered a likely member of the cluster. Due to diffusion {effects that are present in the abundances of	main-sequence and turn-off stars \citep{souto_2018}, and the lack of calibrated DR14 abundances for dwarfs observed in the APOGEE survey \citep[][Holtzman et al. 2018, {\it submitted}]{dr14}, we restricted our final sample to stars having $\log(g) \lessapprox 3.7$.} 
{Stars passing an RV and proper motion membership cut but falling above the $\log(g) \approx 3.7$ cut are identified as dwarf members (``DM")\footnote{We note the existence of a few stars with missing calibrated $\log(g)$ values that consequently fail our $\log(g)$ cut, even though they are likely giant members, which result in a "dwarf member" (DM) classification.}, while those falling below the $\log(g)$ cut are identified as giant members (``GM")}. Only the giant members are included in the final OCCAM sample. {All stars not falling into either the DM or GM category are identified as non-members (``NM").} Table \ref{tab:full_sample_stars} shows the sample of stars used, with the relevant stellar parameters used and final membership determinations. 

The 19 clusters studied in this work were chosen because they had at least 4 member {\em giant} stars.

\subsubsection{Verifying Membership}

Figure \ref{fig:ngc6819}a shows the CMD for NGC 7789, with identified APOGEE members shown in orange and non-members in blue; \ref{fig:ngc6819}c shows the cluster area on the sky for reference. While some members may have been falsely rejected, obvious non-members are clearly rejected. The $T_{eff}-\log(g)$ diagram in Figure \ref{fig:ngc6819}b shows likely members where they are expected. Figure \ref{fig:ngc6819}d shows a proper motion contour plot, from the 2D Gaussian fit discussed above, which shows members where they are expected. Figure \ref{fig:ngc6819}d also shows some proper motion members rejected for RV and/or metallicity. 

All APOGEE cluster and star data, including membership probabilities and abundances plus bulk cluster properties, will be released as part of a SDSS DR14 mini-data release scheduled for the end of July 2018. The catalog will be available at:  \url{http://www.sdss.org/dr14/data_access/value-added-catalogs/}

\begin{deluxetable*}{lrrrrrrrrrc}
\tablecaption{OCCAM Data Sample \label{tab:full_sample}}
	\tablehead{
    \colhead{Cluster} & 
    \colhead{l} &
    \colhead{b} &
    \colhead{Radius\tablenotemark{a}} &
    \colhead{Age\tablenotemark{a}} &
    \colhead{R$_{GC}$\tablenotemark{b}} & 
    \colhead{$\mu_{\alpha}$\tablenotemark{c}} &
    \colhead{$\mu_{\delta}$\tablenotemark{c}} &
    \colhead{RV} & 
    \colhead{[Fe/H]} & 
    \colhead{Member}\\[-2ex] 
    \colhead{name} &
    \colhead{deg} &
    \colhead{deg} &
    \colhead{(\arcmin)} &
    \colhead{Gyr} &
    \colhead{(kpc)} &
    \colhead{(mas yr$^{-1}$)} &
    \colhead{(mas yr$^{-1}$)} &
    \colhead{(km s$^{-1}$)} & 
    \colhead{(dex)} & 
    \colhead{stars} 
    }
	\startdata
NGC 6791            & 69.9658 &  $+$10.9080 & 6.3 & 4.42 & 7.70 & $-$0.42 $\pm$ 0.25 & $-$2.28 $\pm$ 0.29 & $-$47.3 $\pm$ 1.4 & $+$0.42 $\pm$ 0.05 &  31\\
NGC 6819            & 73.9834 &  $+$8.4882 & 6.9 & 1.62 & 7.70 & $-$2.92 $\pm$ 0.18 & $-$3.86 $\pm$ 0.20 & $+$2.4 $\pm$ 1.7 & $+$0.11 $\pm$ 0.03 &  36\\
NGC 6811            & 79.2233 &  $+$12.0047 & 7.2 & 0.64 & 7.87 & $-$3.39 $\pm$ 0.17 & $-$8.78 $\pm$ 0.18 & $+$7.8 $\pm$ 0.3 & $-$0.01 $\pm$ 0.02 &   4\\
Berkeley 53         & 90.3051 &  $+$3.7555 & 7.5 & 1.23 & 8.90 & $-$3.77 $\pm$ 0.39 & $-$5.69 $\pm$ 0.34 & $-$36.3 $\pm$ 0.5 & $-$0.00 $\pm$ 0.02 &   5\\
NGC 7789            & 115.5392 &  $-$5.3644 & 14.4 & 1.84 & 9.13 & $-$0.93 $\pm$ 0.19 & $-$1.93 $\pm$ 0.20 & $-$54.7 $\pm$ 1.3 & $+$0.05 $\pm$ 0.03 &  17\\
FSR 0494            & 120.0882 &  $+$1.0206 & 5.7 & 2.00 & 10.60 & $-$2.45 $\pm$ 0.48 & $-$0.65 $\pm$ 0.48 & $-$63.3 $\pm$ 1.5 & $+$0.01 $\pm$ 0.02 &   5\\
NGC 188             & 122.8416 &  $+$22.3840 & 17.7 & 4.47 & 9.06 & $-$2.31 $\pm$ 0.19 & $-$0.96 $\pm$ 0.16 & $-$41.5 $\pm$ 1.1 & $+$0.14 $\pm$ 0.03 &  13\\
IC 166              & 130.0502 &  $-$0.1616 & 7.5 & 1.00 & 11.47 & $-$1.46 $\pm$ 0.15 & $+$1.13 $\pm$ 0.28 & $-$40.5 $\pm$ 1.5 & $-$0.06 $\pm$ 0.02 &  15\\
Berkeley 66         & 139.4199 &  $+$0.1803 & 3.3 & 1.41 & 11.55 & $-$0.14 $\pm$ 0.61 & $+$0.01 $\pm$ 0.69 & $-$50.1 $\pm$ 0.3 & $-$0.13 $\pm$ 0.02 &   6\\
King 5              & 143.7732 &  $-$4.2760 & 8.4 & 1.23 & 10.01 & $-$0.26 $\pm$ 0.28 & $-$1.16 $\pm$ 0.29 & $-$44.3 $\pm$ 1.5 & $-$0.11 $\pm$ 0.02 &   5\\
NGC 1245            & 146.6533 &  $-$8.9081 & 11.4 & 1.06 & 10.66 & $+$0.52 $\pm$ 0.23 & $-$1.57 $\pm$ 0.19 & $-$29.2 $\pm$ 0.8 & $-$0.06 $\pm$ 0.02 &  23\\
King 7              & 149.7993 &  $-$1.0215 & 11.1 & 0.71 & 10.54 & $+$1.07 $\pm$ 0.55 & $-$1.21 $\pm$ 0.42 & $-$11.9 $\pm$ 2.0 & $-$0.05 $\pm$ 0.02 &   4\\
NGC 1798            & 160.6994 &  $+$4.8502 & 5.4 & 2.00 & 12.50 & $+$0.89 $\pm$ 0.33 & $-$0.33 $\pm$ 0.31 & $+$2.0 $\pm$ 1.7 & $-$0.18 $\pm$ 0.02 &   9\\
Berkeley 17         & 175.6574 &  $-$3.6494 & 7.2 & 3.98 & 11.08 & $+$2.55 $\pm$ 0.41 & $-$0.32 $\pm$ 0.27 & $-$73.4 $\pm$ 0.4 & $-$0.11 $\pm$ 0.03 &   7\\
Berkeley 71         & 176.6384 &  $+$0.8936 & 4.8 & 1.05 & 11.51 & $+$0.68 $\pm$ 0.36 & $-$1.62 $\pm$ 0.46 & $-$8.7 $\pm$ 2.3 & $-$0.20 $\pm$ 0.03 &   7\\
Teutsch 51          & 182.7401 &  $+$0.4760 & 2.7 & 0.53 & 11.68 & $+$0.56 $\pm$ 0.29 & $-$0.34 $\pm$ 0.34 & $+$17.0 $\pm$ 1.4 & $-$0.28 $\pm$ 0.03 &   5\\
NGC 2158            & 186.6394 &  $+$1.7807 & 8.4 & 2.14 & 12.41 & $-$0.18 $\pm$ 0.32 & $-$2.01 $\pm$ 0.25 & $+$27.5 $\pm$ 1.5 & $-$0.15 $\pm$ 0.03 &  18\\
NGC 2420            & 198.1134 &  $+$19.6318 & 7.5 & 2.32 & 10.25 & $-$1.19 $\pm$ 0.22 & $-$2.13 $\pm$ 0.18 & $+$74.2 $\pm$ 0.5 & $-$0.12 $\pm$ 0.02 &  15\\
NGC 2682            & 215.6906 &  $+$31.9221 & 33.0 & 3.43 & 8.60 & $-$10.97 $\pm$ 0.24 & $-$2.95 $\pm$ 0.24 & $+$33.8 $\pm$ 1.0 & $+$0.07 $\pm$ 0.03 &  35\\
\enddata
\tablenotetext{a}{From \citet{mwsc_catalog}}
\tablenotetext{b}{Calculating using \cite{bailer-jones} with a solar R$_{GC} = 8$ kpc.}
\tablenotetext{c}{$\mu_{\alpha}$ and $\mu_{\delta}$ and their $1 \sigma $ uncertainties are those of the 2D Gaussian fit, discussed in \S\ref{sec:mem_criteria}.}
\tablenotetext{d}{A separate analysis of APOGEE data for the open cluster IC 166 found 13 member stars, and an average [Fe/H] of $-0.08 \pm 0.05$. These results can be found in Sciappacasse-Ulloa et al. (2018, {\it submitted})}
\end{deluxetable*}

\subsection{Measured Cluster Bulk Abundances}

A ``high reliability" criterion is adopted for a cluster to be included in our sample: 4 or more likely member stars, as determined above. This resulted in a total sample of 259 member stars in 19 clusters used for the analysis of galactic abundance gradients. 

The final value for [Fe/H] used for computing metallicity gradients is taken to be the mean metallicity of the likely members.
The uncertainty on this value is taken to be the standard deviation of the mean metallicity for the cluster. 
{We note that the uncertainties in the metallicities for the individual stars as reported in DR14 are typically $\sim 0.01$ dex, which may be an underestimation. We therefore disregard these uncertainties in our consideration of the uncertainty in the cluster metallicity. We find the majority of clusters have an uncertainty of 0.02-0.03 dex, with the exception of one cluster (King 7) which has a standard deviation of only 0.01; we therefore enforce a more conservative 0.02 dex uncertainty for this cluster.}


Our final sample, assuming a solar distance to the Galactic center of 8 kpc and using the {median distance to likely members (stellar distances are taken from \citealt{bailer-jones}; this is discussed in detail in \S\ref{sec:dist}) } is presented in Table \ref{tab:full_sample}. 

\section{Cluster Metallicities in Comparison to previous work}

\begin{figure*}
	\epsscale{1.2}
	\plotone{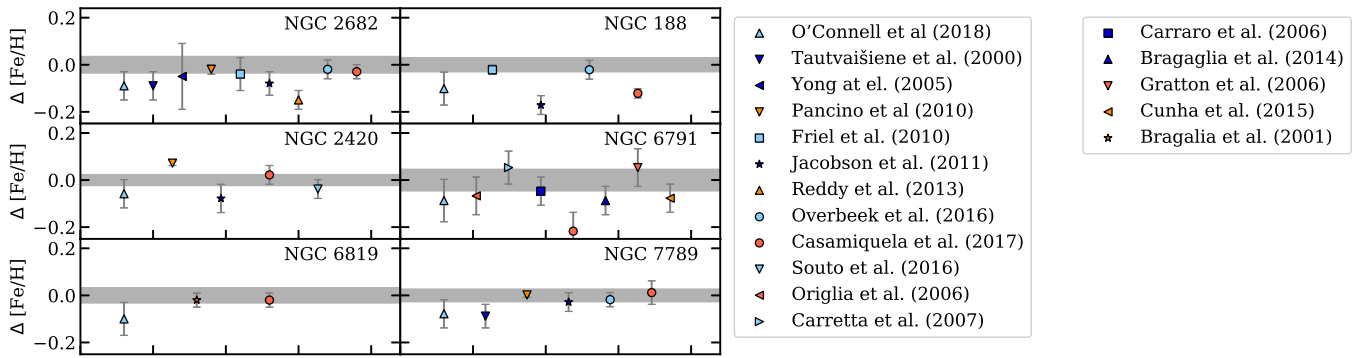}
	\caption{A comparison to commonly studied clusters in the literature ($\Delta$[Fe/H] = literature--OCCAM). The gray bar indicates the internal 1$\sigma$ standard deviation for stars in each cluster from our data. 
	}
	\label{fig:lit_comp}
\end{figure*}

{In order to place our results in the context of previous work, we conducted a detailed comparison of key well-studied clusters from the literature: NGC 188, NGC 2682 (M67), NGC 2420, NGC 6791, NGC 6819, and NGC 7789}, presented in Figure \ref{fig:lit_comp} and discussed below.

\subsection{NGC 2682 (M67)}

Figure \ref{fig:lit_comp} shows that all of the literature values \citep[O'Connell et al.\ 2018;][]{M67_Taut2000,yong_2005,pancino_2010,friel2010,jacobson_2011,reddy_13,Casamiquela_2017} for M67 agree within quoted uncertainties, except for the lowest metallicity value from \cite{reddy_13}, in which the authors note a possible metal-poor offset from the literature. The mean difference from the literature values is {$-0.06 \pm 0.04$} dex for this cluster.

\subsection{NGC 188}
We find that three studies \citep[O'Connell et al.\ 2018;][]{friel2010,overbeek_2016} are in agreement with our results, however we find significant differences with \citet{jacobson_2011} and \citet{Casamiquela_2017}.
The mean difference from the literature is {$-0.09 \pm 0.06$} dex for this cluster, the highest of the commonly studied clusters analyzed.

\subsection{NGC 6791}\label{6791vlit}
We find general agreement with all but one of the literature values considered \citep[][finds a significantly lower metallicity]{Casamiquela_2017}, and note the majority of literature values  \citep[O'Connell et al.\ 2018;][]{origlia_2006,Carretta_2007,carraro_2006,bragaglia_2014,cunha_15} again fall slightly below ours, with a mean difference from the literature of {$-0.06 \pm 0.08$}  dex.
We hypothesize that this may be due to a poor calibration in the metal-rich end of the APOGEE calibrations, as \citet{cunha_15} using APOGEE spectra of many of the same stars for a detailed individual analysis found a lower metallicity ([Fe/H] = 0.34 $\pm$ 0.06) as compared with our DR14 pipeline value ([Fe/H] = 0.42 $\pm$ 0.05). We note, however, that \citet{cunha_15} used an older version of the ASPCAP line list and the DR14 ASPCAP results, and that the \citet{cunha_15} results agree within the uncertainties given the changes in the line list.

\subsection{NGC 2420, NGC 6819, and NGC 7789}


Nearly all of the literature results  are in good agreement with ours, with the exception of \citealt{pancino_2010} for NGC 2420, which quotes particularly small errors. The other results \citep[O'Connell et al.\ 2018;][]{jacobson_2011,Casamiquela_2017} are consistent, {and we note in particular the agreement with \citealt{souto_2016}, who completed a by-hand analysis of the same APOGEE spectra}. The mean differences from the literature are $-0.02 \pm 0.05$, $-0.05 \pm 0.04$, and $-0.03 \pm 0.04$ for NGC 2420, NGC 6819 \citep[O'Connell et al.\ 2018;][]{bragaglia_2001,Casamiquela_2017}, and NGC 7789 \citep[O'Connell et al.\ 2018;][]{M67_Taut2000,pancino_2010,jacobson_2011,overbeek_2016,Casamiquela_2017} respectively.

\subsection{APOGEE DR14 vs. Literature Trends}

The comparison for NGC 188, NGC 6791, and NGC 2682 clearly shows the majority of literature values are more metal poor than our adopted values. The other clusters agree on the direction of the offset, but suggest it is not severe.
J\"onson et al. (2018, {\it submitted}) {compared to optical studies for 525 stars in common with APOGEE.
They find an average difference (in the sense APOGEE -- literature) of $-0.04 \pm 0.010$ dex. From our comparison of 6 open clusters to 17 studies in the literature (Figure \ref{fig:lit_comp}),  we find a mean difference (in the sense APOGEE -- literature) of $-0.05 \pm 0.06$.} 
{Both our analysis and the analysis of J\"onson et al. (2018, {\it submitted}) suggest the possibility of a {\em slight} global metal-rich offset in the APOGEE DR14 sample, but both analyses are consistent with no offset from the literature.}
Still: we emphasize that when using only the APOGEE DR14 sample for analysis, any global offset, minor or otherwise, will have no significant effect on the results of a {\em gradient} measurement.



\section{Galactic Metallicity Gradients} \label{sec:discussion}

The uniform OCCAM sample of 259 member stars in 19 open clusters was used to measure the Galactic metallicity gradient. 
The sample covers the disc from R$_{GC}\approx 7$ to 13 kpc with no major gaps.

In addition to uniform abundances, a uniform $R_{GC}$ analysis is desirable. {We considered 4 sources for cluster distances, discussed in detail in \S\ref{sec:dist}. We use distances computed from the Bailer-Jones Catalog \citep{bailer-jones} for the gradients we present below.}

\begin{figure}
	\centering
    \epsscale{1.2}
    \plotone{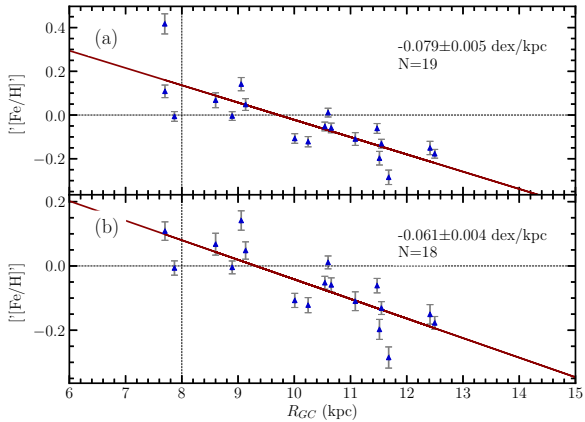}
	\caption{The high reliability metallicity gradients using APOGEE clusters. Dotted lines are shown for reference at $R_{\odot} = 8$ kpc and [Fe/H] = 0 dex. (a) shows the entire sample and (b) shows the sample with the very metal-rich   NGC 6791 removed.
	\label{fig:gradients}}
\end{figure}

\subsection{Error Analysis}\label{sec:error}

The scatter in the abundance gradients necessitates some reliable determination of the uncertainty in any quoted gradient. A re-sampling routine is used to estimate the error in gradients. The gradient is determined {2000} times, each time using a randomly determined [Fe/H] for each cluster, sampled from a standard normal distribution within their uncertainties. The mean of the resulting sample of {2000} gradients is adopted, and the standard deviation is taken as the uncertainty. A check was made against the $\chi^2$ minimum error for a straight line fit in each case; it was found that in every case, our error estimation was larger than this minimum. These are the uncertainties quoted for all the gradients we present.

\subsection{NGC 6791 and the metallicity gradient}

The overall metallicity gradient with the entire sample included, is found to be {$-0.079 \pm 0.005$} dex kpc$^{-1}$ (Figure \ref{fig:gradients}a). We note that NGC 6791 is very metal rich, fairly old, and relatively far from the Galactic plane. Previous work using APOGEE data has suggested it likely migrated to its current location \citep{linden_2017}. Since it is likely not representative of the region of the Galaxy in which it currently resides, we exclude it from further analysis. 
{Previous work including NGC 6791 such as \citet{carraro_98, friel_02}, used a much lower value for [Fe/H] (+0.19 dex and +0.11 dex respectively), low enough to be in disagreement with most recent studies. Even R16 used a lower value for NGC 6791 (+0.24 dex). \citet{jacobson_2011} note that it strongly influences the gradient.}

Removing NGC 6791 gives a final metallicity gradient, from the full OCCAM high-reliability sample, of {$-0.061 \pm 0.004$} dex kpc$^{-1}$ (Figure \ref{fig:gradients}b).

\begin{deluxetable*}{lllll}
	\tablecaption{A summary of reported spectroscopic metallicity gradients. The number of clusters studied by the authors is given, as well as the total number of clusters (including those drawn from the literature) used for the measurement. The range of R$_{GC}$ covered is given, as well as the point at which the authors split their two-function gradient (if any). 
Studies which included only very young or very old clusters are excluded, as were studies that covered a significantly different range in R$_{GC}$.
	\label{tab:grads}}
	\tablehead{\colhead{Study} & \colhead{dex kpc$^{-1}$} & \colhead{\#  study} & \colhead{\# total } & \colhead{range } \\
    \colhead{} & \colhead{} & \colhead{} & \colhead{} & \colhead{kpc} }
    \decimals
	\startdata
	\cite{carraro_98}  $\phantom{jkljkljkljkljkl}$  & $-$0.085  $\pm$ 0.008$\phantom{jkl}$ & $\phn$0 $\phantom{jkl}$ & 37$\phantom{jkl}$ & 7--16 $\phantom{jkl}$\\
	\cite{friel_02}      & $-$0.06$\phn$  $\pm$ 0.01    & 24      & 39 & 7--16\\
    \cite{carrera_2011} \tablenotemark{a}  & $-$0.070  $\pm$ 0.010    & $\phn$9 & 89& 6--12.5 \\
	\cite{jacobson_2011} & $-$0.085 $\pm$ 0.019   & 10      & 19 & 9--13  \\
	\cite{yong_2012} \tablenotemark{a}     & $-$0.09$\phn$  $\pm$ 0.01    & $\phn$5 & 49 & 6--13 \\
	\cite{reddy_16}\tablenotemark{a}    & $-$0.052 $\pm$ 0.011   & 28      & 79 & 5--12 \\
    This Study   & $-$0.061 $\pm$ 0.004   & 18     & 18 & 7--12  \\
	\enddata
    \tablenotetext{a}{These studies fit a two-function gradient. We quote only the gradient measured for the inner sample, as we only discuss this measurement.}
\end{deluxetable*}

\subsection{Distance Effects on the Gradient}\label{sec:dist}

\begin{figure}
	\centering
	\epsscale{1.2}
	\plotone{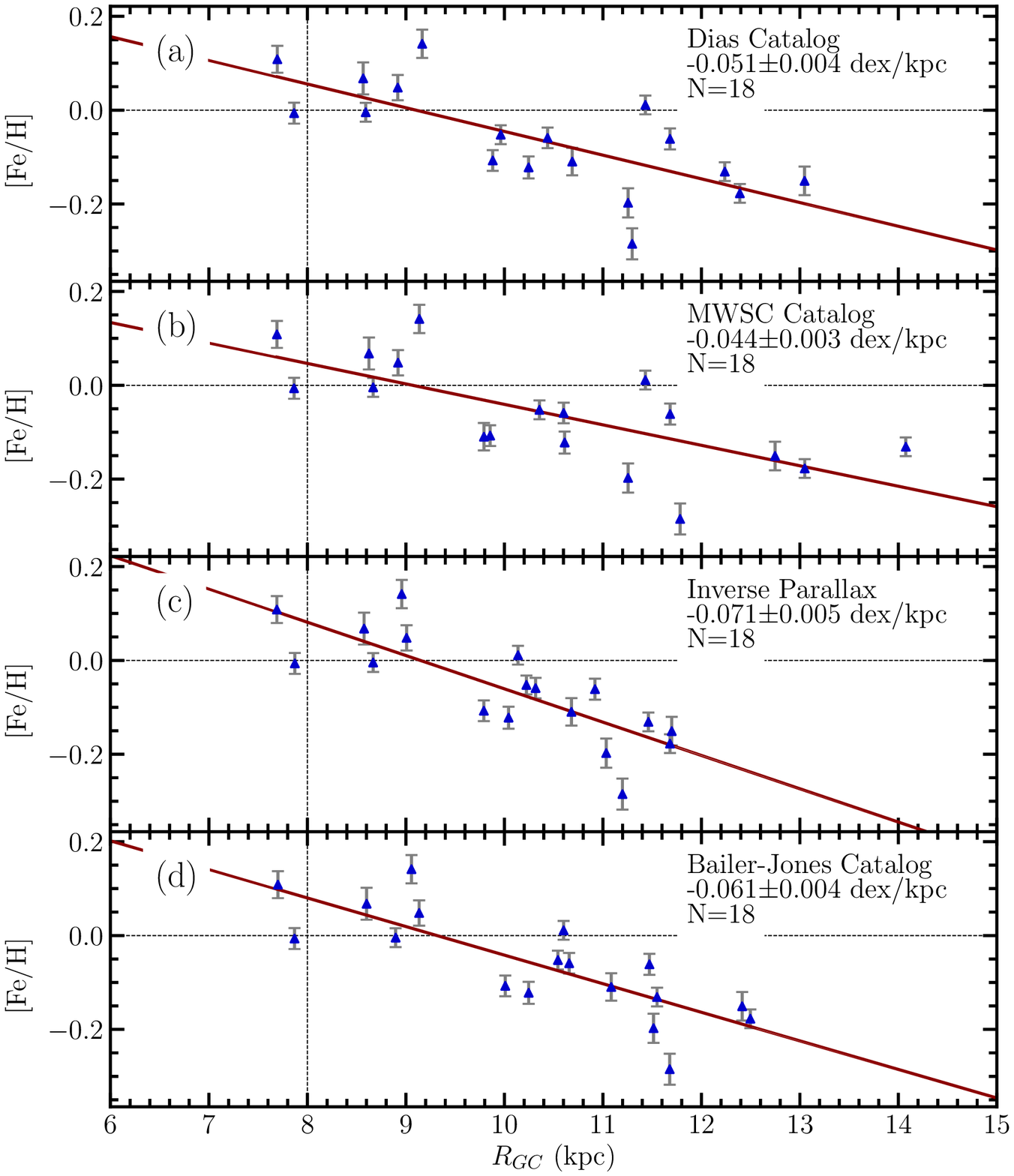}
	\caption{The galactic metallicity gradient computed using values from the Dias Catalog \citep{dias_catalog}, the MWSC Catalog \citep{mwsc_catalog}, inverse-parallax \citep{gaia_dr2, gaia_astrometry}, and the Bailer-Jones catalog \citep{bailer-jones}.}
	\label{fig:distance}
\end{figure}

{The metallicity gradients are highly susceptible to systematic differences in the distance values used. We considered 4 sources of distances: 1) the Dias Catalog \citep{dias_catalog}, which is a compilation of distances from the literature, 2) the MWSC Catalog \citep{mwsc_catalog}, which recomputed distances to each cluster, 3) inverse-parallax \citep{gaia_dr2, gaia_astrometry}, accounting for a 0.08 milli-arcsecond offset \citep{parallax_offset}, and 4) the Bailer-Jones catalog \citep{bailer-jones}, which used the same parallax measurements combined with a geometric prior to compute distances to nearly every star in \gaia DR2. Table \ref{tab:distances} shows a summary of R$_{GC}$ using these 4 sources for distance, while Figure \ref{fig:distance} shows the $d$[Fe/H]/$d$R$_{GC}$ gradient computed using different distance catalogs/methods. 
	
Not surprisingly, the parallax and Bailer-Jones distances are fairly similar, at least for relatively nearby clusters, yet still: the gradient measurements are incompatible. The Dias Catalog and MWSC Catalog distances are similar for a number of clusters, but for clusters at higher R$_{GC}$ they tend to be larger than Bailer-Jones or parallax, leading to a shallower gradient result for both catalogs. The Dias Catalog and MWSC Catalog gradients are barely in agreement within the uncertainties. 
	
The MWSC Catalog recomputed distances for every cluster, and thus is internally consistent. The parallax distances and Bailer-Jones distances are internally consistent as well. But there is a clear discrepancy between these 3 data sets. The Bailer-Jones geometric distances should be the most accurate for nearby clusters, since they are based on \gaia parallaxes. Considering clusters within 7 kpc $<$ R$_{GC}$ $<$ 10 kpc, the MWSC Catalog distances are in good agreement with Bailer-Jones distances. For some more distant clusters, significant discrepancies exist (e.g Berkeley 66 and Berkeley 17), while many remain in good agreement. At this time, we have no strong evidence to distrust one catalog over the other at larger distances, but a decision must be made. Looking at the 2 clusters with very discrepant MWSC Catalog distances (Berkeley 66 and Berkeley 17), we see they also disagree significantly with the Dias Catalog, so for this study, we adopt distances from Bailer-Jones.
} 

\begin{deluxetable}{lrccc}
	\tabletypesize{\small}
	\tablecaption{R$_{GC}$ calculated using different distance sources \label{tab:distances}}
	\tablehead{
		\colhead{Cluster} & 
		\colhead{R$_{GC}$} &
		\colhead{R$_{GC}$} &
		\colhead{R$_{GC}$} &
		\colhead{R$_{GC}$}   
		\\[-1ex]
		\colhead{name} & 
		\colhead{Dias} &
		\colhead{MWSC} &
		\colhead{Parallax}&
		\colhead{Bailer-Jones}   
		\\[-1ex]
		\colhead{} &
		\colhead{(kpc)} &
		\colhead{(kpc)} &
		\colhead{(kpc)} &
		\colhead{(kpc)} 
	}
	\startdata
	NGC 6791       &  7.83 &   7.80 &  7.57 &  7.70\\
	NGC 6819       &  7.69 &   7.69 &  7.69 &  7.70\\
	NGC 6811       &  7.86 &   7.86 &  7.87 &  7.87\\
	Berkeley 53    &  8.59 &   8.67 &  8.67 &  8.90\\
	NGC 7789       &  8.92 &   8.92 &  9.01 &  9.13\\
	FSR 0494       & 11.43 &  11.43 & 10.14 & 10.60\\
	NGC 188        &  9.17 &   9.14 &  8.96 &  9.06\\
	IC 166         & 11.68 &  11.68 & 10.92 & 11.47\\
	Berkeley 66    & 12.24 &  14.08 & 11.46 & 11.55\\
	King 5         &  9.88 &   9.86 &  9.79 & 10.01\\
	NGC 1245       & 10.44 &  10.60 & 10.32 & 10.66\\
	King 7         &  9.96 &  10.36 & 10.22 & 10.54\\
	NGC 1798       & 12.39 &  13.05 & 11.68 & 12.50\\
	Berkeley 17    & 10.69 &   9.79 & 10.68 & 11.08\\
	Berkeley 71    & 11.26 &  11.26 & 11.03 & 11.51\\
	Teutsch 51     & 11.30 &  11.78 & 11.20 & 11.68\\
	NGC 2158       & 13.05 &  12.75 & 11.70 & 12.41\\
	NGC 2420       & 10.25 &  10.61 & 10.04 & 10.25\\
	NGC 2682       &  8.57 &   8.62 &  8.58 &  8.60
	\enddata
\end{deluxetable}
\clearpage
\subsection{Comparison to Previous Work}

A summary of current results in the literature {(from studies using high-resolution spectroscopy)} is found in Table \ref{tab:grads}. We omit studies that measure a gradient in a region significantly different than that considered in this paper. We can readily compare the APOGEE metallicity gradients to these results.

\begin{deluxetable*}{lrrrrrrrrrrr}
\tabletypesize{\tiny}
\tablecaption{OCCAM DR14 Cluster Abundances \label{tab:other_elem}}
	\tablehead{
    \colhead{Cluster} & 
    \colhead{[O/Fe]} &
    \colhead{[Mg/Fe]} &
    \colhead{[Si/Fe]} &
    \colhead{[S/Fe]} &
    \colhead{[Ca/Fe]} &
    \colhead{[V/Fe]} &
    \colhead{[Cr/Fe]} &
    \colhead{[Mn/Fe]} &
    \colhead{[Co/Fe]} &
    \colhead{[Ni/Fe]}   
    \\[-4ex]
    \colhead{name} &
    \colhead{(dex)} &
    \colhead{(dex)} &
    \colhead{(dex)} &
    \colhead{(dex)} &
    \colhead{(dex)} &
    \colhead{(dex)} &
    \colhead{(dex)} &
    \colhead{(dex)} &
    \colhead{(dex)} &
    \colhead{(dex)} &
 }
	\startdata
NGC 6791            & 0.07 $\pm$ 0.04 & 0.06 $\pm$ 0.06 & -0.01 $\pm$ 0.05 &  0.05 $\pm$ 0.11 & 0.02 $\pm$ 0.06 & -0.01 $\pm$ 0.16& -0.11 $\pm$ 0.08 & -0.00 $\pm$ 0.14 & 0.04 $\pm$ 0.27 & -0.00 $\pm$ 0.04\\
NGC 6819            & -0.02 $\pm$ 0.03 & 0.00 $\pm$ 0.02 & 0.00 $\pm$ 0.03 &  -0.02 $\pm$ 0.05 & 0.01 $\pm$ 0.02 & 0.02 $\pm$ 0.07& 0.02 $\pm$ 0.03 & 0.03 $\pm$ 0.02 & 0.05 $\pm$ 0.08 & 0.02 $\pm$ 0.02\\
NGC 6811            & -0.09 $\pm$ 0.04 & -0.02 $\pm$ 0.02 & 0.00 $\pm$ 0.03 &  0.05 $\pm$ 0.05 & -0.00 $\pm$ 0.03 & -0.05 $\pm$ 0.08& 0.01 $\pm$ 0.04 & -0.00 $\pm$ 0.02 & -0.21 $\pm$ 0.12 & -0.02 $\pm$ 0.02\\
Berkeley 53         & -0.02 $\pm$ 0.03 & -0.02 $\pm$ 0.02 & 0.01 $\pm$ 0.03 &  0.03 $\pm$ 0.06 & 0.01 $\pm$ 0.03 & 0.00 $\pm$ 0.08& -0.03 $\pm$ 0.04 & -0.01 $\pm$ 0.03 & -0.31 $\pm$ 0.30 & -0.02 $\pm$ 0.02\\
NGC 7789            & -0.03 $\pm$ 0.03 & -0.02 $\pm$ 0.02 & -0.02 $\pm$ 0.02 &  -0.00 $\pm$ 0.05 & -0.02 $\pm$ 0.02 & -0.01 $\pm$ 0.09& 0.00 $\pm$ 0.05 & -0.01 $\pm$ 0.02 & -0.07 $\pm$ 0.09 & -0.03 $\pm$ 0.02\\
FSR 0494            & -0.05 $\pm$ 0.05 & -0.04 $\pm$ 0.02 & -0.02 $\pm$ 0.03 &  -0.01 $\pm$ 0.08 & -0.00 $\pm$ 0.04 & 0.12 $\pm$ 0.11& 0.03 $\pm$ 0.06 & 0.02 $\pm$ 0.04 & 0.05 $\pm$ 0.22 & 0.01 $\pm$ 0.03\\
NGC 188             & 0.02 $\pm$ 0.04 & 0.05 $\pm$ 0.02 & 0.01 $\pm$ 0.02 &  0.01 $\pm$ 0.08 & -0.02 $\pm$ 0.02 & 0.03 $\pm$ 0.08& -0.01 $\pm$ 0.06 & 0.08 $\pm$ 0.03 & 0.13 $\pm$ 0.11 & 0.04 $\pm$ 0.02\\
IC 166              & -0.02 $\pm$ 0.07 & 0.02 $\pm$ 0.04 & 0.07 $\pm$ 0.06 &  0.04 $\pm$ 0.14 & -0.00 $\pm$ 0.04 & -0.12 $\pm$ 0.27& 0.00 $\pm$ 0.06 & 0.00 $\pm$ 0.04 & -0.41 $\pm$ 0.60 & -0.02 $\pm$ 0.03\\
Berkeley 66         & 0.04 $\pm$ 0.10 & 0.06 $\pm$ 0.03 & 0.05 $\pm$ 0.03 &  0.01 $\pm$ 0.07 & -0.00 $\pm$ 0.04 & -0.14 $\pm$ 0.17& 0.01 $\pm$ 0.06 & -0.05 $\pm$ 0.04 & -0.01 $\pm$ 0.22 & -0.03 $\pm$ 0.05\\
King 5              & 0.00 $\pm$ 0.04 & -0.02 $\pm$ 0.02 & 0.03 $\pm$ 0.03 &  0.07 $\pm$ 0.06 & -0.00 $\pm$ 0.03 & -0.01 $\pm$ 0.11& 0.04 $\pm$ 0.04 & -0.03 $\pm$ 0.04 & -0.00 $\pm$ 0.12 & -0.01 $\pm$ 0.02\\
NGC 1245            & -0.03 $\pm$ 0.07 & -0.03 $\pm$ 0.02 & 0.03 $\pm$ 0.03 &  0.00 $\pm$ 0.07 & -0.00 $\pm$ 0.03 & 0.04 $\pm$ 0.09& 0.01 $\pm$ 0.05 & -0.01 $\pm$ 0.03 & -0.17 $\pm$ 0.50 & -0.04 $\pm$ 0.02\\
King 7              & -0.02 $\pm$ 0.03 & -0.01 $\pm$ 0.02 & 0.04 $\pm$ 0.03 &  0.10 $\pm$ 0.07 & 0.00 $\pm$ 0.02 & -0.06 $\pm$ 0.07& 0.03 $\pm$ 0.03 & 0.03 $\pm$ 0.03 & -0.04 $\pm$ 0.07 & -0.05 $\pm$ 0.02\\
NGC 1798            & 0.01 $\pm$ 0.06 & -0.01 $\pm$ 0.02 & 0.01 $\pm$ 0.03 &  0.01 $\pm$ 0.06 & 0.03 $\pm$ 0.03 & -0.03 $\pm$ 0.11& -0.00 $\pm$ 0.05 & -0.07 $\pm$ 0.03 & -0.18 $\pm$ 0.25 & -0.03 $\pm$ 0.02\\
Berkeley 17         & 0.03 $\pm$ 0.03 & 0.06 $\pm$ 0.02 & 0.02 $\pm$ 0.03 &  0.06 $\pm$ 0.05 & 0.01 $\pm$ 0.03 & 0.02 $\pm$ 0.07& 0.03 $\pm$ 0.04 & -0.01 $\pm$ 0.03 & 0.06 $\pm$ 0.07 & 0.03 $\pm$ 0.02\\
Berkeley 71         & 0.02 $\pm$ 0.09 & 0.03 $\pm$ 0.05 & 0.06 $\pm$ 0.03 &  0.13 $\pm$ 0.08 & 0.03 $\pm$ 0.03 & 0.03 $\pm$ 0.11& 0.01 $\pm$ 0.07 & -0.04 $\pm$ 0.04 & -0.14 $\pm$ 0.23 & -0.03 $\pm$ 0.02\\
Teutsch 51          & 0.08 $\pm$ 0.08 & 0.01 $\pm$ 0.03 & 0.06 $\pm$ 0.06 &  0.01 $\pm$ 0.13 & 0.03 $\pm$ 0.05 & -0.00 $\pm$ 0.13& 0.05 $\pm$ 0.08 & -0.03 $\pm$ 0.05 & 0.09 $\pm$ 0.21 & -0.01 $\pm$ 0.04\\
NGC 2158            & 0.00 $\pm$ 0.07 & 0.03 $\pm$ 0.02 & 0.03 $\pm$ 0.03 &  0.10 $\pm$ 0.09 & -0.00 $\pm$ 0.03 & -0.15 $\pm$ 0.13& -0.07 $\pm$ 0.12 & -0.05 $\pm$ 0.04 & -0.07 $\pm$ 0.22 & -0.01 $\pm$ 0.03\\
NGC 2420            & 0.05 $\pm$ 0.06 & 0.00 $\pm$ 0.03 & 0.01 $\pm$ 0.03 &  -0.01 $\pm$ 0.06 & 0.03 $\pm$ 0.03 & -0.09 $\pm$ 0.13& -0.04 $\pm$ 0.10 & -0.04 $\pm$ 0.03 & -0.18 $\pm$ 0.20 & -0.02 $\pm$ 0.02\\
NGC 2682            & -0.03 $\pm$ 0.04 & 0.01 $\pm$ 0.02 & -0.03 $\pm$ 0.02 &  -0.02 $\pm$ 0.05 & -0.02 $\pm$ 0.02 & -0.08 $\pm$ 0.13& -0.02 $\pm$ 0.07 & 0.01 $\pm$ 0.02 & -0.00 $\pm$ 0.09 & 0.02 $\pm$ 0.02\\
\enddata
\end{deluxetable*}

\begin{figure*}[t!]
	\centering
	\epsscale{1.2}
	\plotone{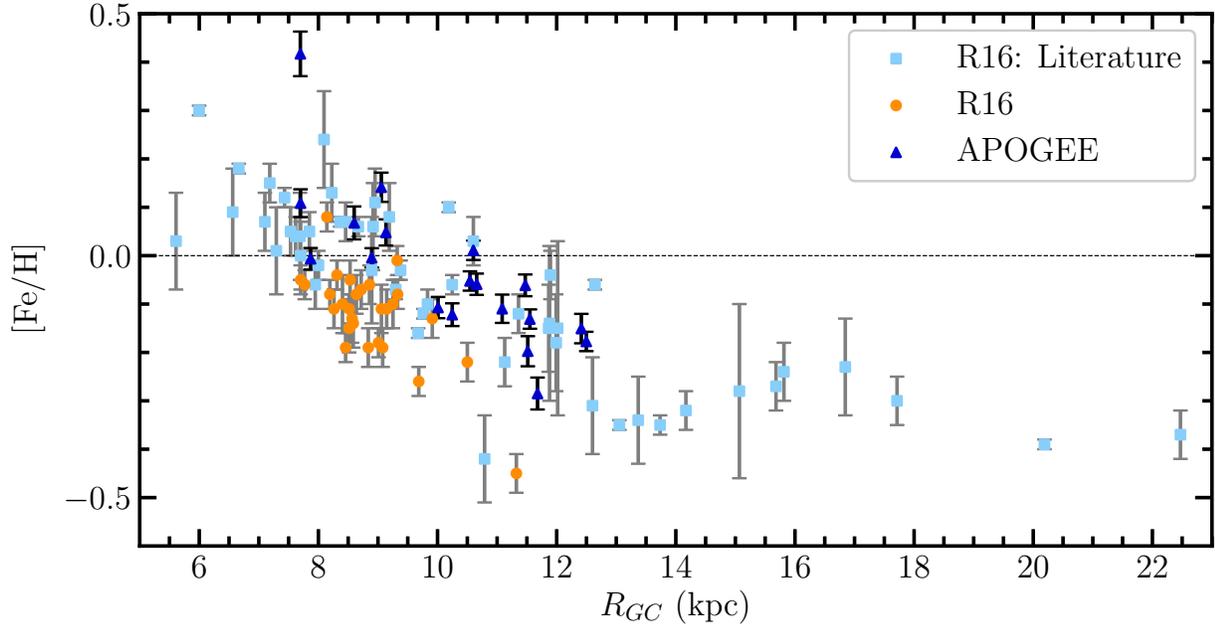}
	\caption{The Galactic abundance trend (R$_{GC}$ vs.[Fe/H]) assuming R$_{GC,\sun}$ = 8.0 kpc.  Our sample (dark blue triangles) is shown along with a literature sample from \citet{reddy_16} (light blue points), and clusters analyzed by \cite{reddy_12, reddy_13, reddy_15, reddy_16} (orange points). 
	}
	\label{fig:full_sample}
\end{figure*}

\subsubsection{Comparison to APOGEE DR12}

{We recomputed the metallicity gradient found from APOGEE DR12 \citep{cunha_16_grads} using only clusters in common with this work (and excluding NGC 6791) and distances from \citet{dias_catalog} and found a gradient of $-$0.035 $\pm$ 0.014. This agrees within the uncertainties with our gradient measured using the Dias catalog. The differences between the metallicity gradients can be explained in terms of improvements in the data reduction of APOGEE spectra, line list, and methodology (Holtzman et al. 2018, {\it submitted}).}
	

\subsubsection{Comparison to the Other Work}

{We find a metallicity gradient consistent with 4 of the 6 studies, the 2 discrepent results being \cite{carraro_98} and \cite{yong_2012}, which both quote particularly steep gradients. We find a relatively close agreement with \cite{friel_02}, \cite{carrera_2011}, and R16. We note that if we instead compare the metallicity gradient computed with NGC 6791, our result is in agreement with \cite{carraro_98} and \cite{yong_2012}, but would no longer be in agreement with R16 or \cite{friel_02}.}
It is worth emphasizing that R16 and \citet{friel_02} both had large uniform samples (24 and 28 open clusters, respectively) in addition to the literature samples included in their studies. Since the R16 study is both recent and very large, we compare to it directly. In Figure \ref{fig:full_sample} we show
the sample uniformly analyzed in R16 (orange points) and their literature compiled sample (light blue points), along with the APOGEE results (dark blue triangles).

\section{Other Elements Beyond [Fe/H]}

{We compute } mean DR14 cluster abundances for reliable $\alpha$-related elements  (O, 
Ca, Mg, Si, S) and iron peak elements (Cr, Co, Ni, Mn, V)  {in the same manner as [Fe/H]}, shown in Table \ref{tab:other_elem}. The abundances are shown for individual stars in Table \ref{cl_metals}. 


\begin{deluxetable*}{lrrccrrrr}[h!]
\tabletypesize{\tiny}
\tablecaption{DR14 OCAAM Open Cluster Member Star Abundances \tablenotemark{a} \label{cl_metals}}
\tablehead{ 
\colhead{Cluster}  & 
  \colhead{2MASS ID} &
  \colhead{RV} 
 &\colhead{[Fe/H]}&\colhead{[O/Fe]}
&\colhead{[Mg/Fe]}&\colhead{[Ca/Fe]} 
&\colhead{[Si/Fe]}&\colhead{[S/Fe]} 
\\[-4ex] 
& \colhead{} 
&  \colhead{(km s$^{-1}$)} 
& \colhead{(dex)}& \colhead{(dex)}
& \colhead{(dex)}& \colhead{(dex)}
& \colhead{(dex)}& \colhead{(dex)}\\\\[-4ex] 
&&&&\colhead{[V/Fe]}&\colhead{[Cr/Fe]} 
&\colhead{[Mn/Fe]}&\colhead{[Co/Fe]} 
&\colhead{[Ni/Fe]}  \\[-4ex] 
&&&&\colhead{(dex)}& \colhead{(dex)}
& \colhead{(dex)}& \colhead{(dex)}
& \colhead{(dex)} 
}
\startdata
NGC 6819 & 2M19404803+4008085 & 2.4 $\pm$ 0.1 & 0.09 $\pm$ 0.01 & -0.01 $\pm$ 0.02 & 0.02 $\pm$ 0.02 & 0.01 $\pm$ 0.02 &  0.02 $\pm$ 0.02 &-0.00 $\pm$ 0.05
 \\ &&&& 0.04 $\pm$ 0.06 & 0.02 $\pm$ 0.03 & 0.01 $\pm$ 0.02 & 0.07 $\pm$ 0.06 & 0.01 $\pm$ 0.01  \\ [1ex] 
NGC 6819 & 2M19404965+4014313 & 3.1 $\pm$ 0.0 & 0.12 $\pm$ 0.01 & 0.01 $\pm$ 0.03 & 0.01 $\pm$ 0.02 & 0.02 $\pm$ 0.02 &  0.00 $\pm$ 0.02 &-0.03 $\pm$ 0.05
 \\ &&&& 0.06 $\pm$ 0.07 & 0.02 $\pm$ 0.04 & 0.03 $\pm$ 0.03 & 0.07 $\pm$ 0.08 & 0.02 $\pm$ 0.02  \\ [1ex] 
NGC 6819 & 2M19405020+4013109 & 4.3 $\pm$ 0.0 & 0.15 $\pm$ 0.01 & 0.00 $\pm$ 0.03 & 0.00 $\pm$ 0.02 & -0.00 $\pm$ 0.02 &  -0.04 $\pm$ 0.02 &-0.04 $\pm$ 0.05
 \\ &&&& -0.02 $\pm$ 0.06 & -0.02 $\pm$ 0.03 & 0.01 $\pm$ 0.02 & 0.03 $\pm$ 0.07 & 0.02 $\pm$ 0.01  \\ [1ex] 
NGC 6819 & 2M19405601+4013395 & 3.3 $\pm$ 0.1 & 0.09 $\pm$ 0.01 & 0.04 $\pm$ 0.04 & -0.00 $\pm$ 0.02 & 0.02 $\pm$ 0.03 &  -0.13 $\pm$ 0.03 &-0.07 $\pm$ 0.06
 \\ &&&& -0.01 $\pm$ 0.09 & -0.00 $\pm$ 0.05 & 0.02 $\pm$ 0.03 & 0.10 $\pm$ 0.10 & 0.00 $\pm$ 0.02  \\ [1ex] 
NGC 6819 & 2M19405797+4008174 & 4.5 $\pm$ 0.1 & 0.13 $\pm$ 0.01 & 0.00 $\pm$ 0.03 & 0.02 $\pm$ 0.02 & 0.00 $\pm$ 0.02 &  0.01 $\pm$ 0.02 &-0.06 $\pm$ 0.05
 \\ &&&& 0.08 $\pm$ 0.07 & 0.03 $\pm$ 0.03 & 0.05 $\pm$ 0.02 & 0.03 $\pm$ 0.07 & 0.00 $\pm$ 0.01  \\ [1ex] 
NGC 6819 & 2M19410524+4014042 & 3.3 $\pm$ 0.1 & 0.14 $\pm$ 0.01 & -0.03 $\pm$ 0.03 & -0.01 $\pm$ 0.02 & 0.04 $\pm$ 0.02 &  0.02 $\pm$ 0.03 &0.03 $\pm$ 0.05
 \\ &&&& 0.02 $\pm$ 0.07 & 0.02 $\pm$ 0.04 & 0.03 $\pm$ 0.03 & -0.04 $\pm$ 0.08 & 0.04 $\pm$ 0.02  \\ [1ex] 
NGC 6819 & 2M19410622+4010532 & 3.2 $\pm$ 0.0 & 0.12 $\pm$ 0.01 & 0.07 $\pm$ 0.04 & 0.01 $\pm$ 0.02 & 0.02 $\pm$ 0.03 &  0.03 $\pm$ 0.03 &-0.05 $\pm$ 0.06
 \\ &&&& 0.02 $\pm$ 0.08 & -0.04 $\pm$ 0.04 & 0.02 $\pm$ 0.03 & 0.03 $\pm$ 0.10 & 0.02 $\pm$ 0.02  \\ [1ex] 
NGC 6819 & 2M19410858+4013299 & 2.3 $\pm$ 0.0 & 0.11 $\pm$ 0.01 & -0.06 $\pm$ 0.03 & 0.00 $\pm$ 0.02 & -0.01 $\pm$ 0.02 &  0.03 $\pm$ 0.02 &-0.03 $\pm$ 0.05
 \\ &&&& 0.03 $\pm$ 0.06 & 0.03 $\pm$ 0.03 & 0.03 $\pm$ 0.02 & 0.08 $\pm$ 0.07 & 0.01 $\pm$ 0.01  \\ [1ex] 
NGC 6819 & 2M19410926+4014436 & 2.3 $\pm$ 0.1 & 0.13 $\pm$ 0.01 & -0.03 $\pm$ 0.03 & 0.00 $\pm$ 0.02 & -0.00 $\pm$ 0.02 &  -0.03 $\pm$ 0.02 &-0.02 $\pm$ 0.05
 \\ &&&& -0.04 $\pm$ 0.06 & -0.00 $\pm$ 0.03 & 0.00 $\pm$ 0.02 & 0.01 $\pm$ 0.07 & -0.02 $\pm$ 0.01  \\ [1ex] 
NGC 6819 & 2M19410991+4015495 & 2.5 $\pm$ 0.1 & 0.07 $\pm$ 0.01 & -0.03 $\pm$ 0.03 & -0.00 $\pm$ 0.02 & 0.05 $\pm$ 0.02 &  0.01 $\pm$ 0.03 &0.02 $\pm$ 0.05
 \\ &&&& 0.13 $\pm$ 0.07 & 0.06 $\pm$ 0.04 & 0.00 $\pm$ 0.03 & 0.13 $\pm$ 0.08 & 0.03 $\pm$ 0.02  \\ [1ex]  
\enddata
\tablenotetext{a}{Table \ref{cl_metals} is published in its entirety in the
  electronic edition of Astronomical Journal. A portion is shown here for guidance regarding its form and content.}
\end{deluxetable*}

\subsection{OCCAM DR14 Calibration Sample}

\begin{figure*}
\begin{center}
\epsscale{1.15}
\plotone{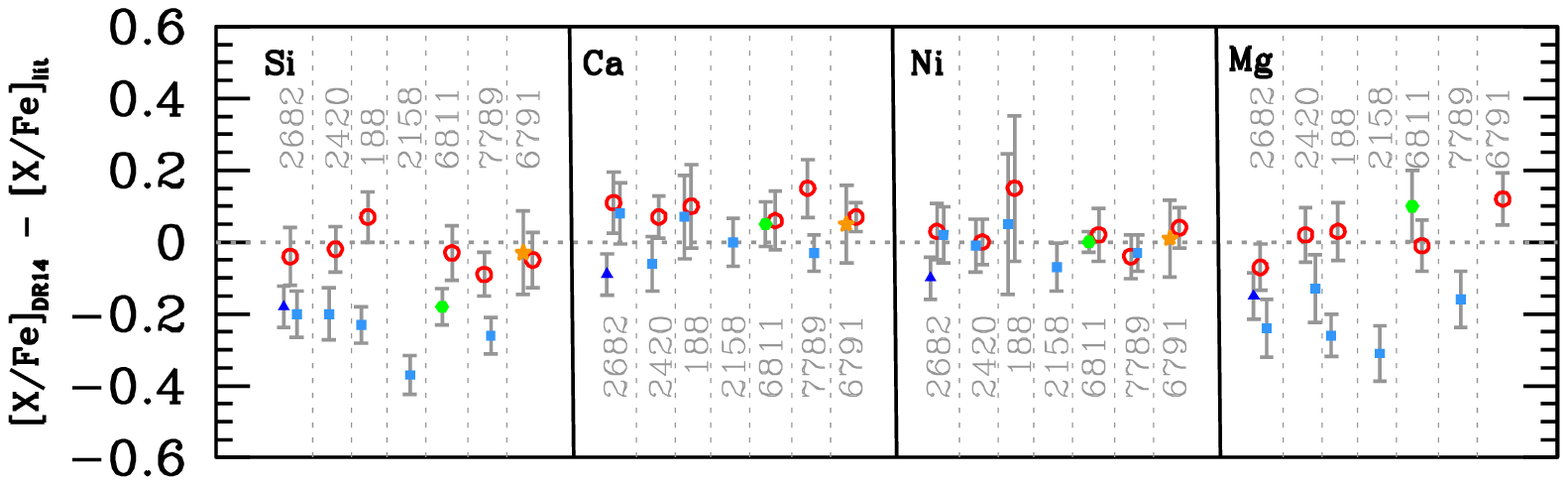}
\end{center}
\caption{ \label{compSBSfeh}
Comparison of individual elemental abundances using the APOGEE calibration clusters
with available comparison elements in the other literature studies.  
Clusters are color-coded by analysis group: 
  {\it dark blue} for \citealt{reddy_13,reddy_16},
  {\it light blue} for \citealt{jacobson_2011},
  {\it green} for \citealt{bragaglia_2001},
  {\it orange} for \citealt{carraro_2006}, and
  {\it red} for O'Connell et al. 2018 (submitted).
}
\end{figure*}

We use the APOGEE calibration cluster set to search for systematics in other available elements (Si, Ca, Ni, Mg). As shown in Figure~\ref{compSBSfeh} and listed in Table \ref{cl_calib}, there are no significant systematic offsets, with the possible exception of [Mg/Fe] {and [Si/Fe]. For all other elements, the offsets are within the uncertainties (see \S \ref{sec:error}) for nearly every study and cluster}. 
{J\"onson et al. (2018, {\it submitted}) also perform a detailed comparison to the literature for these elements, and find no significant systematic offsets.}
For [Mg/Fe], the APOGEE data are offset from the \citealt{jacobson_2011} clusters, but not from \citealt{bragaglia_2001} and \citealt{carraro_2006} data.
These [Mg/Fe] discrepancies are most likely due to line-list differences and will require further exploration; however, since we are consistent with some clusters and likely the effect is systematic between groups, we apply no offset here.  

\subsection{Galactic Gradients in Other Elements}

The full APOGEE DR14 sample allows an exploration of individual abundance
gradients for key element groups\footnote{While the DR14 APOGEE sample also contains C and N, these elements have strong stellar evolutionary changes along the giant branch.  Given the small numbers of stars per cluster, we have excluded these elements from consideration in this paper.}, such as $\alpha$-related elements  and iron peak elements. 
These elements are key for exploring how Galactic
chemical enrichment occurs, as each element is produced in a different manner (e.g., SNII vs. SNIa yield ratios). 

\begin{figure}
	\epsscale{1.2}
	\plotone{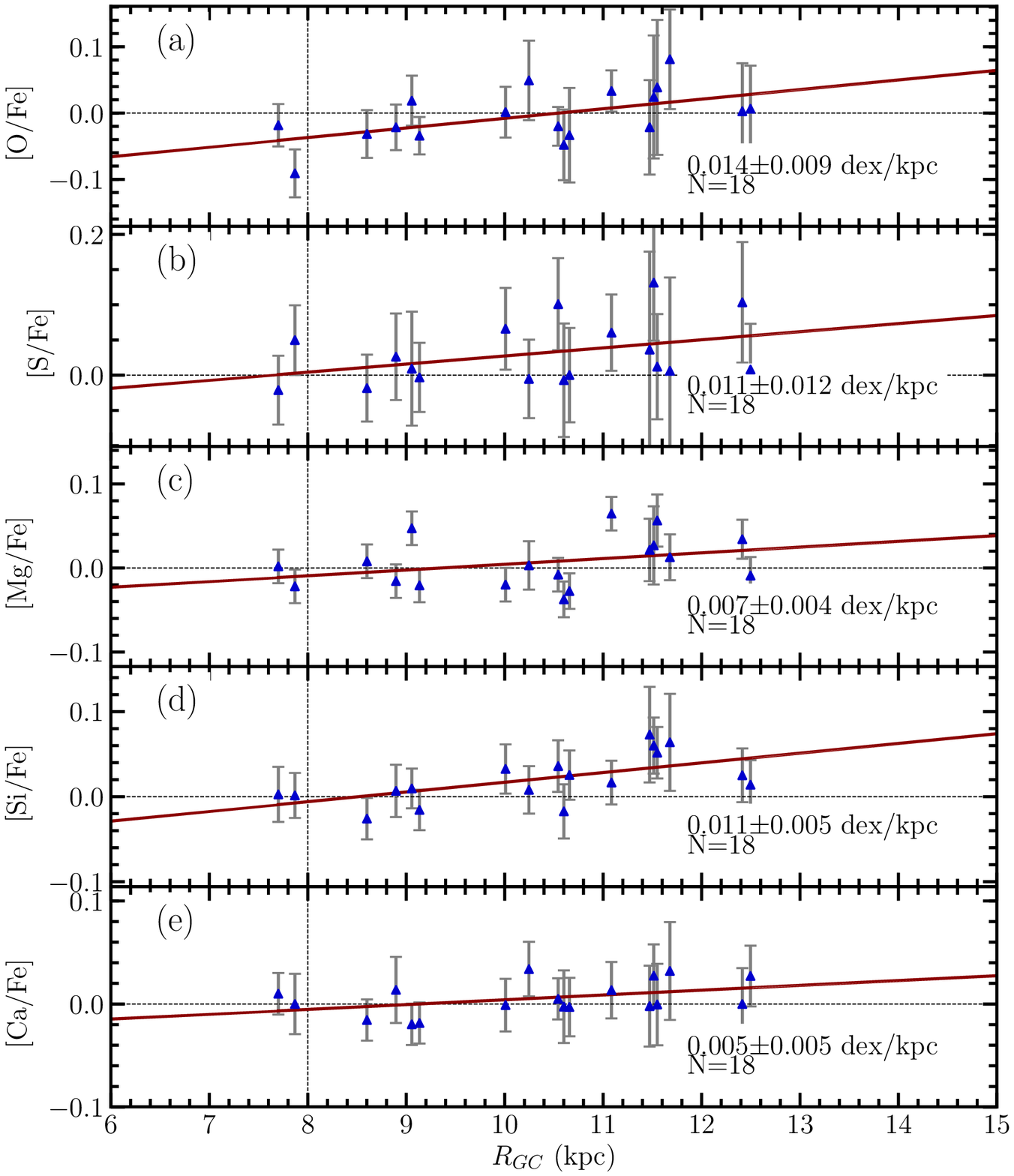}
	\caption{Galactic Trend for our sample for the $\alpha$ elements (O, S, Mg, Si, Ca) from DR14.}
	\label{fig:alpha}
\end{figure}

We find statistically significant {\it increasing} trends for some of the $\alpha$ elements ([O/Fe], [Mg/Fe], and [Si/Fe]),  seen in Figure \ref{fig:alpha}. {The other $\alpha$ trends (Si, Ca) also show a positive trend, but their large uncertainties make them also consistent with a slope of 0}. This behavior is consistent with previous work \citep[e.g., ][]{jacobson_2011}, who also found a significant $d$[O/Fe]/$d$R trend from a literature compiled cluster sample.

This mild positive [$\alpha$/Fe] gradient is in agreement with the chemical evolution models of \citet{minchev_2}, who find an [Mg/Fe] gradient (averaged over all age ranges, for $|Z|<0.25$ kpc) of 0.009 dex kpc$^{-1}$, although the gradient for younger populations (which may better match our relatively young sample) is steeper, e.g. 0.027 dex kpc$^{-1}$ for age $<2$ Gyr. The models of \cite{kubryk_2015} also show a qualitatively similar trend for [O/Fe].

\begin{figure}
	\epsscale{1.2}
	\plotone{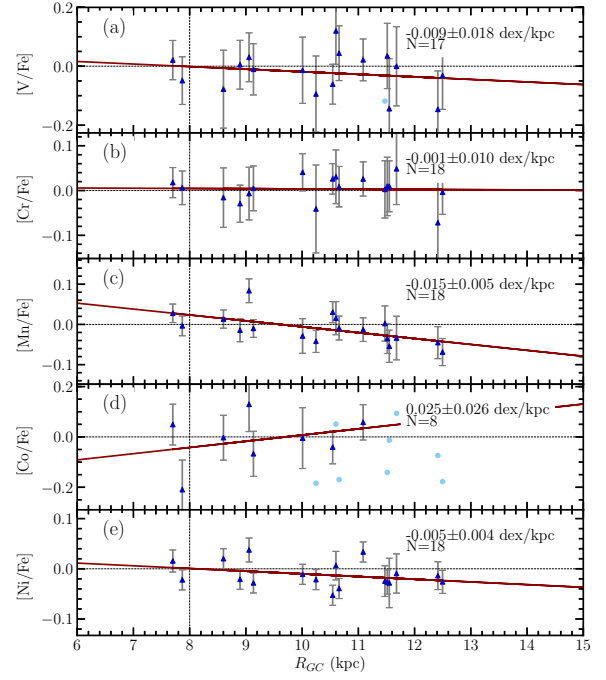}
	\caption{Galactic trend for our sample for the iron-peak (V, Co, Mn, Cr, Ni) elements from DR14. Clusters with very large uncertainties are not included in the fit (N reflects only those included in the fit), but are shown for reference as blue dots.
	}
	\label{fig:iron_peak}
\end{figure}

{We also see a statistically significant {\it decreasing} trend for the iron-peak elements [Mn/Fe] and [Ni/Fe] as seen in Figure \ref{fig:iron_peak}. The uncertainties are too large to draw meaningful conclusions for other elements (V, Cr, Co)}.

\clearpage

\clearpage
\startlongtable
\noindent\begin{deluxetable*}{lrrcrcl}
\tabletypesize{\scriptsize}
\tablecaption{DR14 OCCAM Abundance Comparison to Literature \label{cl_calib}}
\tablehead{ 
\colhead{Cluster}  
&\multicolumn{1}{c}{Abundance}&\multicolumn{1}{l}{\# cluster}&\multicolumn{1}{c}{Abundance}
&\multicolumn{1}{l}{\# cluster} & \colhead{$\Delta$[X/Fe]}
&\multicolumn{1}{c}{References}\\[-4ex] 
&\multicolumn{1}{c}{ Lit.\ (dex)}&\multicolumn{1}{l}{stars
  Lit.}&\multicolumn{1}{c}{DR 14 (dex)} & \multicolumn{1}{l}{stars DR14}
& \colhead{(dex)}&
}
\startdata
\multicolumn{7}{c}{[Si/Fe]}\\\hline
NGC 6819    	&	   $+$0.18 $\pm$   0.04  	&	3	&	  $+$0.00  $\pm$  0.03 	&	36	&	 $-$0.18 	$\pm$ 	0.05	&	 Bragaglia et al.\ 2001 \\
NGC 6819    	&	   $+$0.03 $\pm$   0.07  	&	3	&	  $+$0.00  $\pm$  0.03 	&	36	&	 $-$0.03 	$\pm$ 	0.08	&	 O'Connell et al.\ 2018 \\
NGC 6791    	&	   $+$0.02 $\pm$   0.10  	&	10	&	  $-$0.01  $\pm$  0.06 	&	31	&	 $-$0.03 	$\pm$ 	0.12	&	 Carraro et al.\ 2006 \\
NGC 6791    	&	   $+$0.04 $\pm$   0.05  	&	2	&	  $-$0.01  $\pm$  0.06 	&	31	&	 $-$0.05 	$\pm$ 	0.08	&	 O'Connell et al.\ 2018 \\
NGC 2158    	&	   $+$0.39 $\pm$   0.05  	&	15	&	  $+$0.02  $\pm$  0.02 	&	18	&	 $-$0.37 	$\pm$ 	0.05	&	 Jacobson et al.\ 2009 \\
NGC 188     	&	   $+$0.25 $\pm$   0.05  	&	27	&	  $+$0.02  $\pm$  0.01 	&	13	&	 $-$0.23 	$\pm$ 	0.05	&	 Jacobson et al.\ 2011\\
NGC 188     	&	   $-$0.05 $\pm$   0.07  	&	3	&	  $+$0.02  $\pm$  0.01 	&	13	&	 $+$0.07 	$\pm$ 	0.07	&	 O'Connell et al.\ 2018 \\
NGC 2420    	&	   $+$0.21 $\pm$   0.07  	&	9	&	  $+$0.01  $\pm$  0.02 	&	15	&	 $-$0.20 	$\pm$ 	0.07	&	 Jacobson et al.\ 2011\\
NGC 2420    	&	   $+$0.03 $\pm$   0.06  	&	6	&	  $+$0.01  $\pm$  0.02 	&	15	&	 $-$0.02 	$\pm$ 	0.06	&	 O'Connell et al.\ 2018 \\
NGC 2682    	&	   $+$0.21 $\pm$   0.05  	&	19	&	  $+$0.01  $\pm$  0.04 	&	35	&	 $-$0.20 	$\pm$ 	0.06	&	 Jacobson et al.\ 2011\\
NGC 2682    	&	   $+$0.19 $\pm$   0.04  	&	3	&	  $+$0.01  $\pm$  0.04 	&	35	&	 $-$0.18 	$\pm$ 	0.06	&	 Reddy et al.\ 2013 \\
NGC 2682    	&	   $+$0.05 $\pm$   0.07  	&	10	&	  $+$0.01  $\pm$  0.04 	&	35	&	 $-$0.04 	$\pm$ 	0.08	&	 O'Connell et al.\ 2018 \\
NGC 7789    	&	   $+$0.25 $\pm$   0.05  	&	28	&	  $-$0.01  $\pm$  0.01 	&	17	&	 $-$0.26 	$\pm$ 	0.05	&	 Jacobson et al.\ 2011\\
NGC 7789    	&	   $+$0.08 $\pm$   0.06  	&	5	&	  $-$0.01  $\pm$  0.01 	&	17	&	 $-$0.09 	$\pm$ 	0.06	&	 O'Connell et al.\ 2018 \\\hline
\multicolumn{7}{c}{[Ca/Fe]}\\\hline														
NGC 6819    	&	   $-$0.04 $\pm$   0.06  	&	3	&	  $+$0.01  $\pm$  0.02 	&	36	&	 $+$0.05 	$\pm$ 	0.06	&	 Bragaglia et al.\ 2001 \\
NGC 6819    	&	   $-$0.05 $\pm$   0.08  	&	3	&	  $+$0.01  $\pm$  0.02 	&	36	&	 $+$0.06 	$\pm$ 	0.08	&	 O'Connell et al. 2018 \\
NGC 6791    	&	   $-$0.03 $\pm$   0.10  	&	10	&	  $+$0.02  $\pm$  0.04 	&	31	&	 $+$0.05 	$\pm$ 	0.11	&	 Carraro et al.\ 2006 \\
NGC 6791    	&	   $-$0.05 $\pm$\nodata  	&	2	&	  $+$0.02  $\pm$  0.04 	&	31	&	 $+$0.07 	$\pm$ 	\nodata	&	 O'Connell et al. 2018 \\
NGC 2158    	&	   $+$0.00 $\pm$   0.06  	&	15	&	  $+$0.00  $\pm$  0.03 	&	18	&	 $+$0.00 	$\pm$ 	0.07	&	 Jacobson et al.\ 2009 \\
NGC 188     	&	   $-$0.04 $\pm$   0.06  	&	27	&	  $+$0.03  $\pm$  0.10 	&	13	&	 $+$0.07 	$\pm$ 	0.12	&	 Jacobson et al.\ 2011\\
NGC 188     	&	   $-$0.07 $\pm$   0.06  	&	3	&	  $+$0.03  $\pm$  0.10 	&	13	&	 $+$0.10 	$\pm$ 	0.12	&	 O'Connell et al. 2018 \\
NGC 2420    	&	   $+$0.10 $\pm$   0.07  	&	9	&	  $+$0.04  $\pm$  0.03 	&	15	&	 $-$0.06 	$\pm$ 	0.08	&	 Jacobson et al.\ 2011\\
NGC 2420    	&	   $-$0.03 $\pm$   0.05  	&	6	&	  $+$0.04  $\pm$  0.03 	&	15	&	 $+$0.07 	$\pm$ 	0.06	&	 O'Connell et al. 2018 \\
NGC 2682    	&	   $-$0.11 $\pm$   0.07  	&	19	&	  $-$0.03  $\pm$  0.05 	&	35	&	 $+$0.08 	$\pm$ 	0.09	&	 Jacobson et al.\ 2011\\
NGC 2682    	&	   $+$0.06 $\pm$   0.03  	&	3	&	  $-$0.03  $\pm$  0.05 	&	35	&	 $-$0.09 	$\pm$ 	0.06	&	 Reddy et al. 2013 \\
NGC 2682    	&	   $-$0.14 $\pm$   0.07  	&	10	&	  $-$0.03  $\pm$  0.05 	&	35	&	 $+$0.11 	$\pm$ 	0.09	&	 O'Connell et al. 2018 \\
NGC 7789    	&	   $+$0.01 $\pm$   0.05  	&	28	&	  $-$0.02  $\pm$  0.01 	&	17	&	 $-$0.03 	$\pm$ 	0.05	&	 Jacobson et al.\ 2011\\
NGC 7789    	&	   $-$0.17 $\pm$   0.08  	&	5	&	  $-$0.02  $\pm$  0.01 	&	17	&	 $+$0.15 	$\pm$ 	0.08	&	 O'Connell et al. 2018 \\\hline
\multicolumn{7}{c}{[Ni/Fe]}\\\hline														
NGC 6819    	&	   $+$0.01 $\pm$   0.02  	&	3	&	  $+$0.01  $\pm$  0.02 	&	36	&	 $+$0.00 	$\pm$ 	0.03	&	 Bragaglia et al.\ 2001 \\
NGC 6819    	&	   $-$0.01 $\pm$   0.07  	&	3	&	  $+$0.01  $\pm$  0.02 	&	36	&	 $+$0.02 	$\pm$ 	0.07	&	 O'Connell et al. 2018 \\
NGC 6791    	&	   $-$0.01 $\pm$   0.10  	&	10	&	  $+$0.00  $\pm$  0.04 	&	31	&	 $+$0.01 	$\pm$ 	0.11	&	 Carraro et al.\ 2006 \\
NGC 6791    	&	   $-$0.04 $\pm$   0.04  	&	2	&	  $+$0.00  $\pm$  0.04 	&	31	&	 $+$0.04 	$\pm$ 	0.06	&	 O'Connell et al. 2018 \\
NGC 2158    	&	   $+$0.05 $\pm$   0.06  	&	15	&	  $-$0.02  $\pm$  0.03 	&	18	&	 $-$0.07 	$\pm$ 	0.07	&	 Jacobson et al.\ 2009 \\
NGC 188     	&	   $+$0.08 $\pm$   0.05  	&	27	&	  $+$0.13  $\pm$  0.19 	&	13	&	 $+$0.05 	$\pm$ 	0.20	&	 Jacobson et al.\ 2011\\
NGC 188     	&	   $-$0.02 $\pm$   0.07  	&	3	&	  $+$0.13  $\pm$  0.19 	&	13	&	 $+$0.15 	$\pm$ 	0.20	&	 O'Connell et al. 2018 \\
NGC 2420    	&	   $-$0.01 $\pm$   0.07  	&	9	&	  $-$0.02  $\pm$  0.02 	&	15	&	 $-$0.01 	$\pm$ 	0.07	&	 Jacobson et al.\ 2011\\
NGC 2420    	&	   $-$0.02 $\pm$   0.06  	&	6	&	  $-$0.02  $\pm$  0.02 	&	15	&	 $-$0.00 	$\pm$ 	0.06	&	 O'Connell et al. 2018 \\
NGC 2682    	&	   $-$0.01 $\pm$   0.06  	&	19	&	  $+$0.01  $\pm$  0.05 	&	35	&	 $+$0.02 	$\pm$ 	0.08	&	 Jacobson et al.\ 2011\\
NGC 2682    	&	   $+$0.11 $\pm$   0.03  	&	3	&	  $+$0.01  $\pm$  0.05 	&	35	&	 $-$0.10 	$\pm$ 	0.06	&	 Reddy et al. 2013 \\
NGC 2682    	&	   $-$0.02 $\pm$   0.06  	&	10	&	  $+$0.01  $\pm$  0.05 	&	35	&	 $+$0.03 	$\pm$ 	0.08	&	 O'Connell et al. 2018 \\
NGC 7789    	&	   $+$0.00 $\pm$   0.05  	&	28	&	  $-$0.03  $\pm$  0.01 	&	17	&	 $-$0.03 	$\pm$ 	0.05	&	 Jacobson et al.\ 2011\\
NGC 7789    	&	   $+$0.01 $\pm$   0.06  	&	5	&	  $-$0.03  $\pm$  0.01 	&	17	&	 $-$0.04 	$\pm$ 	0.06	&	 O'Connell et al. 2018 \\\hline
\multicolumn{7}{c}{[Mg/Fe]}\\\hline														
NGC 6819    	&	   $-$0.12 $\pm$   0.07  	&	3	&	  $+$0.00  $\pm$  0.01 	&	36	&	 $+$0.12 	$\pm$ 	0.07	&	 Bragaglia et al.\ 2001 \\
NGC 6819    	&	   $+$0.01 $\pm$   0.07  	&	3	&	  $+$0.00  $\pm$  0.01 	&	36	&	 $-$0.01 	$\pm$ 	0.07	&	 O'Connell et al. 2018 \\
NGC 2158    	&	   $+$0.22 $\pm$   0.07  	&	15	&	  $+$0.03  $\pm$  0.01 	&	18	&	 $-$0.19 	$\pm$ 	0.07	&	 Jacobson et al.\ 2009\\
NGC 188     	&	   $+$0.26 $\pm$   0.05  	&	27	&	  $+$0.03  $\pm$  0.04 	&	13	&	 $-$0.23 	$\pm$ 	0.06	&	 Jacobson et al.\ 2011\\
NGC 188     	&	   $+$0.00 $\pm$   0.07  	&	3	&	  $+$0.03  $\pm$  0.04 	&	13	&	 $+$0.03 	$\pm$ 	0.08	&	 O'Connell et al. 2018 \\
NGC 2420    	&	   $+$0.11 $\pm$   0.09  	&	9	&	  $+$0.00  $\pm$  0.03 	&	15	&	 $-$0.11 	$\pm$ 	0.09	&	 Jacobson et al.\ 2011\\
NGC 2420    	&	   $-$0.02 $\pm$   0.07  	&	6	&	  $+$0.00  $\pm$  0.03 	&	15	&	 $+$0.02 	$\pm$ 	0.08	&	 O'Connell et al. 2018 \\
NGC 2682    	&	   $+$0.23 $\pm$   0.07  	&	19	&	  $-$0.03  $\pm$  0.05 	&	35	&	 $-$0.26 	$\pm$ 	0.09	&	 Jacobson et al.\ 2011\\
NGC 2682    	&	   $+$0.12 $\pm$   0.04  	&	3	&	  $-$0.03  $\pm$  0.05 	&	35	&	 $-$0.16 	$\pm$ 	0.06	&	 Reddy et al. 2013 \\
NGC 2682    	&	   $+$0.04 $\pm$   0.04  	&	10	&	  $-$0.03  $\pm$  0.05 	&	35	&	 $-$0.07 	$\pm$ 	0.06	&	 O'Connell et al. 2018 \\
NGC 7789    	&	   $+$0.14 $\pm$   0.05  	&	28	&	  $-$0.02  $\pm$  0.01 	&	17	&	 $-$0.16 	$\pm$ 	0.05	&	 Jacobson et al.\ 2011\\
NGC 7789    	&	   $-$0.07 $\pm$   0.04  	&	5	&	  $-$0.02  $\pm$  0.01 	&	17	&	 $+$0.05 	$\pm$ 	0.04	&	 O'Connell et al. 2018 
\enddata
\end{deluxetable*}

\section{Conclusions }

{We describe the technique used by the OCCAM survey for targeting likely open cluster members, and another technique for determining the likelihood of their membership in a cluster.}
{Using the determined likely cluster giant members, we present the} first multi-element data from the OCCAM collaboration's exploration of the SDSS/APOGEE open cluster data presented in DR14. 

{We present abundance measurements of 11 elements for 19 open clusters, and find no systematic offsets from previous work in the literature. Using distance measurements from \citet{bailer-jones}, we make measurements of the Galactic abundance gradient for all 11 elements. Our results are in general agreement with previous work, and we present new evidence for a trend in [Mn/Fe] and [Ni/Fe]. Specifically: }
\begin{enumerate}
\item The [Fe/H] gradient is $-0.061 \pm 0.004$ dex kpc$^{-1}$,
  derived from clusters spanning $7 < R_{GC} < 13$ kpc.  
\item We measure a mild [$\alpha$/Fe] gradient,
  including [O/Fe], [Mg/Fe], and [Si/Fe], and mild negative gradient for the iron peak elements [Mn/Fe] and [Ni/Fe]. 
\end{enumerate} 


While our sample of 19 open clusters is one of the largest \textit{uniform} samples to date used to study Galactic abundance gradients, it is still small. {Nevertheless, our results show tight correlations with a linear fit, and good agreement with previous work in the literature, suggesting that APOGEE data can be a powerful tool in studying Galactic abundance gradients.} Future work will feature more clusters from {larger number of open clusters 	observed by APOGEE and comparisons with chemical evolution and chemodynamical models.}

\acknowledgements

JD, PMF, BT, JO, KMJ and BMM acknowledge support for this research from the National Science Foundation (AST-1311835 \& AST-1715662) and the TCU RCAF and JFSRP programs.
KMJ also acknowledges funding from a TCU SERC grant and NSF REU (PHY-0851558). 
KC acknowledges support for this research from the
National Science Foundation (AST-0907873).
DAGH and OZ acknowledge support provided by the Spanish Ministry of
Economy and Competitiveness (MINECO) under grant AYA-2017-88254-P.
We would like to thank J. G. Fern{\'a}ndez-Trincado and Doug Geisler for their helpful comments in the manuscript.
The authors would also like to thank the 
Max-Planck-Institut f\"ur Astronomie (MPIA Heidelberg) for hosting JD and PMF during the completion of this work.

Funding for SDSS-III has been provided by the Alfred P. Sloan Foundation, the Participating Institutions, the National Science Foundation, and the U.S. Department of Energy Office of Science. The SDSS-III web site is http://www.sdss3.org/.

SDSS-III is managed by the Astrophysical Research Consortium for the Participating Institutions of the SDSS-III Collaboration including the University of Arizona, the Brazilian Participation Group, Brookhaven National Laboratory, Carnegie Mellon University, University of Florida, the French Participation Group, the German Participation Group, Harvard University, the Instituto de Astrofisica de Canarias, the Michigan State/Notre Dame/JINA Participation Group, Johns Hopkins University, Lawrence Berkeley National Laboratory, Max Planck Institute for Astrophysics, Max Planck Institute for Extraterrestrial Physics, New Mexico State University, New York University, Ohio State University, Pennsylvania State University, University of Portsmouth, Princeton University, the Spanish Participation Group, University of Tokyo, University of Utah, Vanderbilt University, University of Virginia, University of Washington, and Yale University.

Funding for the Sloan Digital Sky Survey IV has been provided by the Alfred P. Sloan Foundation, the U.S. Department of Energy Office of Science, and the Participating Institutions. SDSS-IV acknowledges
support and resources from the Center for High-Performance Computing at
the University of Utah. The SDSS web site is www.sdss.org.

SDSS-IV is managed by the Astrophysical Research Consortium for the 
Participating Institutions of the SDSS Collaboration including the 
Brazilian Participation Group, the Carnegie Institution for Science, 
Carnegie Mellon University, the Chilean Participation Group, the French Participation Group, Harvard-Smithsonian Center for Astrophysics, 
Instituto de Astrof\'isica de Canarias, The Johns Hopkins University, 
Kavli Institute for the Physics and Mathematics of the Universe (IPMU) / 
University of Tokyo, Lawrence Berkeley National Laboratory, 
Leibniz Institut f\"ur Astrophysik Potsdam (AIP),  
Max-Planck-Institut f\"ur Astronomie (MPIA Heidelberg), 
Max-Planck-Institut f\"ur Astrophysik (MPA Garching), 
Max-Planck-Institut f\"ur Extraterrestrische Physik (MPE), 
National Astronomical Observatories of China, New Mexico State University, 
New York University, University of Notre Dame, 
Observat\'ario Nacional / MCTI, The Ohio State University, 
Pennsylvania State University, Shanghai Astronomical Observatory, 
United Kingdom Participation Group,
Universidad Nacional Aut\'onoma de M\'exico, University of Arizona, 
University of Colorado Boulder, University of Oxford, University of Portsmouth, 
University of Utah, University of Virginia, University of Washington, University of Wisconsin, 
Vanderbilt University, and Yale University.

This work has made use of data from the European Space Agency (ESA) mission {\it Gaia} (\url{https://www.cosmos.esa.int/gaia}), processed by the {\it Gaia} Data Processing and Analysis Consortium (DPAC, \url{https://www.cosmos.esa.int/web/gaia/dpac/consortium}). Funding for the DPAC has been provided by national institutions, in particular the institutions participating in the {\it Gaia} Multilateral Agreement.

This research made use of Astropy, a community-developed core Python package for Astronomy (Astropy Collaboration, 2018).

\vspace{5mm}
\facilities{Sloan (APOGEE), FLWO:2MASS, \gaia}

\software{\href{http://www.astropy.org/}{Astropy}}

\nocite{astropy}
\bibliography{Donor.bib}


\end{document}